\documentclass{article}
\usepackage[utf8]{inputenc}
\usepackage{algorithm}
\usepackage{fancyhdr}
\usepackage{lastpage}
\usepackage{gensymb}
\usepackage{graphicx}
\DeclareGraphicsExtensions{.pdf,.jpg,.jpg}
\usepackage{amsmath}
\usepackage{amssymb}
\def\code#1{\texttt{#1}}
\usepackage{multicol}
\usepackage[labelsep=period]{caption}
\usepackage[labelfont=bf]{caption}
\usepackage{subcaption}
\usepackage{algorithm}
\usepackage{authblk}

\title{Deep learning approach in multi-scale prediction of turbulent mixing-layer}

\author[$\dagger$]{Jinu Lee}
\author[$\dagger$]{Sangseung Lee}
\author[$\dagger$]{Donghyun You}
\affil[$\dagger$]{Department of Mechanical Engineering, Pohang University of Science and Technology, 77 Cheongam-ro, Namgu, Pohang, Gyeongbuk 37673, Republic of Korea}

\date{}
\begin{document}
\maketitle
\noindent\textbf{Abstract.}\\
Achievement of solutions in Navier-Stokes equation is one of challenging quests, especially for its closure problem. For achievement of particular solutions, there are variety of numerical simulations including Direct Numerical Simulation (DNS) or Large Eddy Simulation (LES). These methods analyze flow physics through efficient reduced-order modeling such as proper orthogonal decomposition or Koopman method, showing prominent fidelity in fluid dynamics. Generative adversarial network (GAN) is a reprint of neurons in brain as combinations of linear operations, using competition between generator and discriminator. Current paper propose deep learning network for prediction of small-scale movements with large-scale inspections only, using GAN. Therefore DNS result of three-dimensional mixing-layer was filtered blurring out the small-scaled structures, then is predicted of its detailed structures, utilizing Generative Adversarial Network (GAN). This enables multi-resolution analysis being asked to predict fine-resolution solution with only inspection of blurry one. Within the grid scale, current paper present deep learning approach of modeling small scale features in turbulent flow. The presented method is expected to have its novelty in utilization of unprocessed simulation data, achievement of 3D structures in prediction by processing 3D convolutions, and predicting precise solution with less computational costs.

\begin{multicols}{2}
\section{Introduction}
Physical coherence in turbulence is studied through variety of simulations and experiments \cite{tennekes1972first,dimotakis1976mixing}. For projection of nonlinear system into high dimensional linear system, mathematical decomposition for reduced order modeling is widely investigated, including proper orthogonal decomposition\cite{lumley1981coherent}, dynamic mode decomposition\cite{schmid2010dynamic}, Koopman method\cite{koopman1931hamiltonian}.\\
\indent Along with acceleration in CPU- and GPU-s computational power, deep learning is achieving its successes in our near past and present\cite{lecun2015deep}. Deep learning for numerical investigation on partial differential equations also shows wide varieties of applications, including Navier-Stokes equation\cite{miyanawala2017efficient}, turbulence modeling\cite{zhang2015machine} and control\cite{lee1997application}. Motivated from previous version by Lee and You\cite{lee2017prediction, lee2018data}, we predicted small-scaled vortex generation in turbulent mixing-layer flow using generative adversarial network with three-dimensional convolution.\\
\indent Proposed by Goodfellow \textit{et al.}\cite{goodfellow2014generative}, Generative Adversarial Network (GAN) has been showing good performance since 2014. In order to generate unsteady flow features, we decided to adopt video prediction model by Mathieu \textit{et.al.}\cite{mathieu2015deep}. Therefore the model is different from conventional GANs in that there exist a rigid ground truth. Gaussian-blurred DNS data of previous time-steps were provided to our GAN model as condition. Therefore with only inspections to those blurry flow field of past, we obtained large and small eddies of future time-step.\\
\indent We used 3D convolution filter to train flow simulation data of location, velocity, and pressure. Despite non-uniformity in DNS data, no data preprocess was done except for normalization. We provided location of each cells as input to overcome non-uniformity. Total seven parameters were put into the network as input; three Cartesian location, according velocities, and pressure.\\
\indent Thus main purpose of our research is to open new possibilities in precise prediction and the real-time inference\cite{lee2018data}, in approach to the solution of Navier-Stokes equation and its closure problem. Current paper introduced previously conducted approaches the successful one and the failures as well. After the primative trials, we decided utilizing raw simulation data without interpolation or plot should be done. There were two incompatible difficulties of directly using DNS result for deep learning are 1) non-uniformity in cell-centered data distribution, and 2) heavy computational cost per each DNS result.

\section{Direct Numerical Simulation\\of Turbulent 3D Mixing-layer}\label{DNS}
\indent Current research mainly aims on the prediction of vorticies generated in mixing layer, using deep learning approach. In deep learning method, which actually have quite a lot of relevance with the actual learing process in human beings, one of the most important part is \textit{how accurate are the training sets}. In other words, preparing the precise flow simulation data is the observable task and actually is the basis of current research.\\
\indent Several flow simulation methods including direct numerical simulation and large eddy simulation were reviewed by Rogallo and Moin\cite{rogallo1984numerical}. Although previous direct numerical simulation (DNS) was mostly limited in the range of low-Reynolds-number turbulence\cite{spalart1988direct}, nowadays DNS is expanding its compatibility in various flow simulations, with ceaseless evolution in computing power.\\
\indent In current paper as well, with intension to prepare unsteady turbulent mixing layer data, direct numerical simulation had been conducted. With several trials in preparation of mesh grid and the boundary conditions, the simulation data is on its refinement, expectedly with allowance in better understanding of phenomenology of turbulence. Grid convergence test was conducted also, providing the robustness in current research data. More details about numerical simulation will be specified in section \textbf{\ref{simulation}}.
\subsection{Governing Equation}\label{navierstokes}
The following form of Navier-Stokes equation illustrates unsteady incompressible viscous Newtonian fluid, Reynolds-averaged:
\begin{equation}\label{continuity}
\frac{\partial \bar{u_j}}{\partial x_j} = 0
\end{equation}
\begin{equation}\label{ns_equation}
\frac{\partial \bar{u_i}}{\partial t} + \frac{\partial \bar{u_i}\bar{u_j}}{\partial x_j} = -\rho^{-1}\frac{\partial \bar{p}}{\partial x_i} + \nu\frac{\partial^2 \bar{u_i}}{\partial x_j\partial x_j} - \frac{\partial \overline{u_i'u_j'}}{\partial x_j}
\end{equation}
\indent Whole simulations conducted in current paper are non-dimensionalzed by this momentum thickness, including Reynolds number Re$_m=\overline{U}\theta_m^0/\nu$ and the computational domain. $\overline{U}$ denotes average streamwise inflow velocity $(U_1+U_2)/2$ and $\nu$ is kinematic viscosity.

\subsection{Numerical Simulation}\label{simulation}
For presented simulation of spatially growing mixing shear layer, our in-house code$^*$ was used both for solving and generating computational grids. For the solver, our second-order Runge-Kutta scheme was used, in Fortran, multi-processed. Further analyses and discussions will be drawn in section \textbf{\ref{g_conv}}, especially with results of Stanley and Sarkar\cite{stanley1997simulations}'s using fourth-order low-storage Runge-Kutta scheme of Carpenter and Kennedy\cite{carpenter1994fourth}. Courant-Friedrichs-Lewy (CFL) number was set static as 1.0 in every solvers herein. Simulation results were exported in timestep $\Delta t$ such that satisfies 
\begin{equation}
C = \Delta t \frac{\overline{U}}{\theta_m^0}.
\end{equation}
It is kept $C=0.72$ as static constant throughout current simulations. Mach number is not considered, since current simulation is of incompressible fluid flow. \\

\noindent \ref{simulation}.1 \textit{Boundary conditions}\\
\indent One of the most important and the typical point that takes simulation apart from experiment is boundary conditions. There are several obstacles to make flow simulation closer to the real physical domain, regarding discretizing continuous system and boundary shear layers. Setting up completely isolated observation without physical disturbances is highly complicated task, with sufficient experiences.\\
\indent Boundary conditions in current paper can be classified into cartesian coordinates: inflow in $-x$ and outflow in $+x$ direction (streamwise), free in/outflow\cite{stanley1997simulations} condition was given in $\pm y$ direction (transverse), and periodic condition in $\pm z$ direction (spanwise). Here, the word \textit{free in/outflow} as boundary condition can be somewhat vague and not so specified. Stanley and Sarkar used similar condition, described as `Non-reflecting inflow/outflow' boundary condition in \cite{stanley1997simulations}, which will be specifically discussed later in this section.\\
\indent Inflow velocity profile is given as following in steamwise direction only:
\begin{equation}\label{inflow}
u = \overline{U} + \frac{\Delta U}{2}\tanh\frac{y}{2\theta_m^0}
\end{equation}
The initial velocity profile is based on tangent hyperbolic function, where $\overline{U}$ is the average velocity of two streams and $\Delta U = U_1 - U_2 > 0$ the velocity difference. Random number was generated for perturbation, with maximum amount as 10\% of initial velocity average $(U_1+U_2)/2$ at the initialization. $U_1$ and $U_2$ denotes initial velocity given as the ratio $\eta_v = (U_1-U_2)/(U_1+U_2) = 0.33$, where $U_1>U_2$.\\
\indent In current paper, lengthscale is non-dimensionalized by $\theta_m$, the momentum thickness, which is defined as followings.
\begin{equation}
\theta_m = \int_{0}^{\infty}\frac{\rho(y)u(y)}{\rho_0u_0}(1-\frac{u(y)}{u_0}) dy
\end{equation}
Here, since we have investigated incompressible flow, density term can simply be cancelled out. Another famous lengthscale vorticity thickness is linearly dependent with $\theta_m$ in most of free shear-layers\cite{moser1993three}. Since defined
\begin{equation}
\delta_\omega = \frac{\Delta U}{\lceil\frac{\partial U}{\partial y}\rceil}
\end{equation}
, it could easily be achieved in the initial boundary condition condition
\begin{equation}
\frac{\partial U}{\partial y} = \frac{\Delta U}{2}\frac{\partial}{\partial y}\tanh\frac{y}{2\theta_m^0}
\end{equation}
simply yielding $\delta_\omega^0 = 4\theta_m^0$ the relation between momentum thickness and the vorticity thickness at inflow. This could vary according to the input velocity profile, $\delta_\omega = 4.44\theta_m^0$ if Blasius profile is given, or even larger in simulation by Rogers and Moser\cite{rogers1994direct}.\\
\indent With expression of momentum thickness and the corresponding Reynolds number, Kolmogorov lengthscale have following relation:
\begin{equation}
\frac{\eta}{\theta_m^0} = \text{Re}^{-3/4}_m.
\end{equation}
Equivalently the momentum thickness $\theta_m^0$ is $\text{Re}_m^{3/4}$ times of $\eta$. It is definite that $\eta=(\nu^3/\epsilon)^{1/4}$ decreases when kinematic viscosity $\nu$ decrease, or kinetic energy dissipation rate $\epsilon$ increase. Therefore it is said that the smallest scale of turbulent flow is yielded by viscosity domination and/or kinetic energy dissipation into heat energy. See section \textbf{\ref{kolmo}} with \cite{tennekes1972first,james1989lecture} for further explanations on Kolmogorov microscale.\\
\indent Periodic boundary conditions were given for spanwise direction so that plane $z=0$ is identical to the plane $z=L_z$. Following the claim by Riley and Metcalfe\cite{riley1980direct}; the periodicity length was given larger than the spatial scale, in order to suppress the influence of periodicity in current simulation.\\
\indent It is important to note that the real simulated domain is much larger than those analyzed herein, in transverse $y$ direction. In other words, \textit{free in/outflow} condition is achieved, by truncating result in `box', and the \textit{full} result is of slip condition in $\pm y$ direction. The boundary conditions used in \cite{stanley1997simulations} is nonreflecting boundary condition by Thompson\cite{thompson1987time}, with inflow conditions by Poinsot and Lele\cite{poinsot1992boundary}. The referenced paper\cite{stanley1997simulations} is non-reflecting since Stanley and Sarkar simulated compressible flow field.\\
\indent Current paper, which is of \textit{incompressible} as explained in section \textbf{\ref{navierstokes}}, does not consider wave reflection. Therefore actual domain which is taken into account is sufficiently smaller in $y$ direction, than those boundaries of slip condition - equivalently allowing fluid particles to freely maneuver the actual boundary. In other words, slip condition was given sufficiently far away from the actual boundary.\\
\begin{figure}[H]
	\centering
	\includegraphics[scale=0.35]{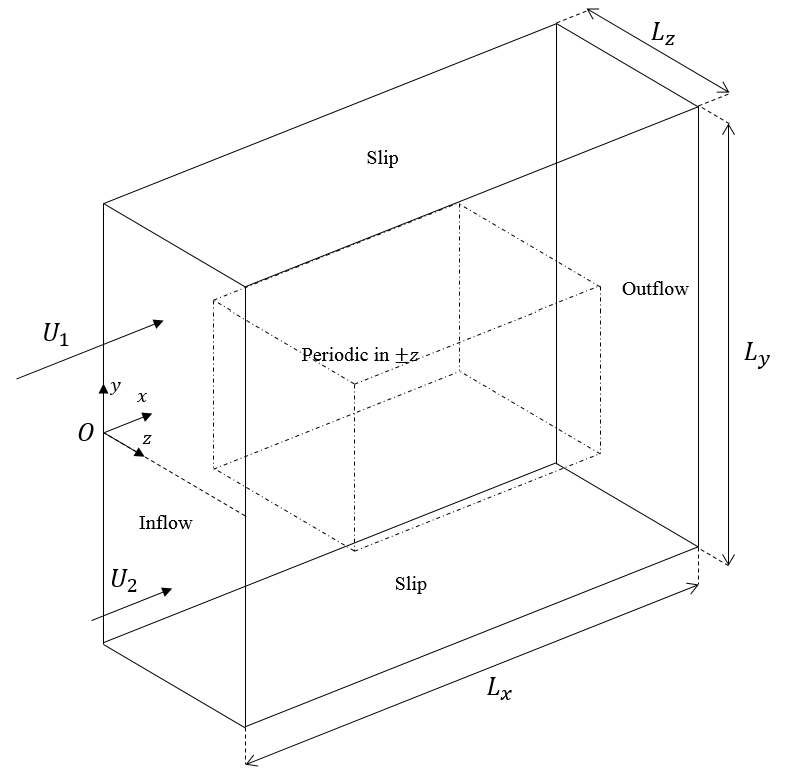}
	\caption{Boundary condition specifications of computational domain. Inflow is given using tangent hyperbolic profile as equation (\ref{inflow}), which is schematically represented in figure. Outer box with solid line indicates whole computational domain $\mathbb{C}$ and inner dotted box represents $\mathbb{B}$ the truncated one.}
	\label{fig:bc}
\end{figure}

\begin{figure}[H]
	\begin{subfigure}{.52\textwidth}
		\centering
		\includegraphics[scale=0.145]{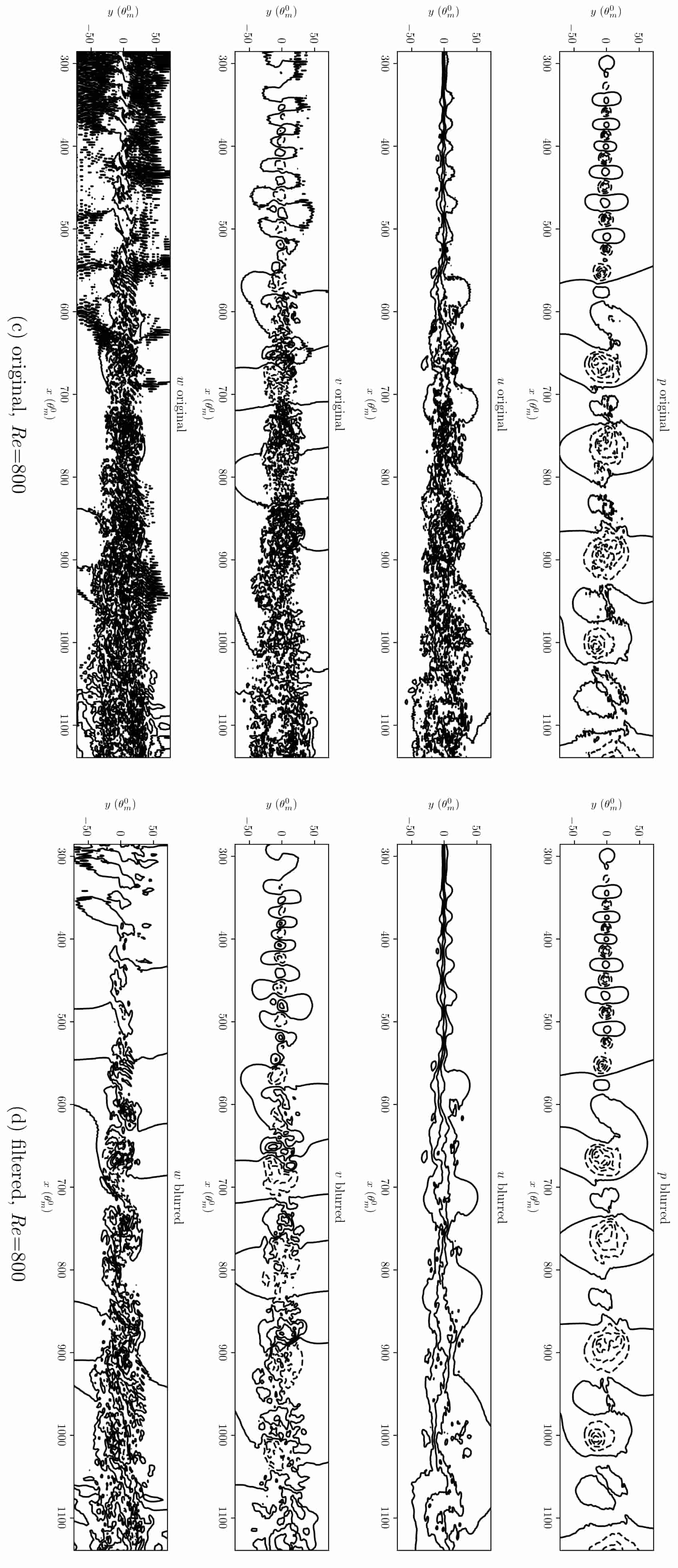}
	\end{subfigure}
	\begin{subfigure}{.52\textwidth}
		\centering
		\includegraphics[scale=0.145]{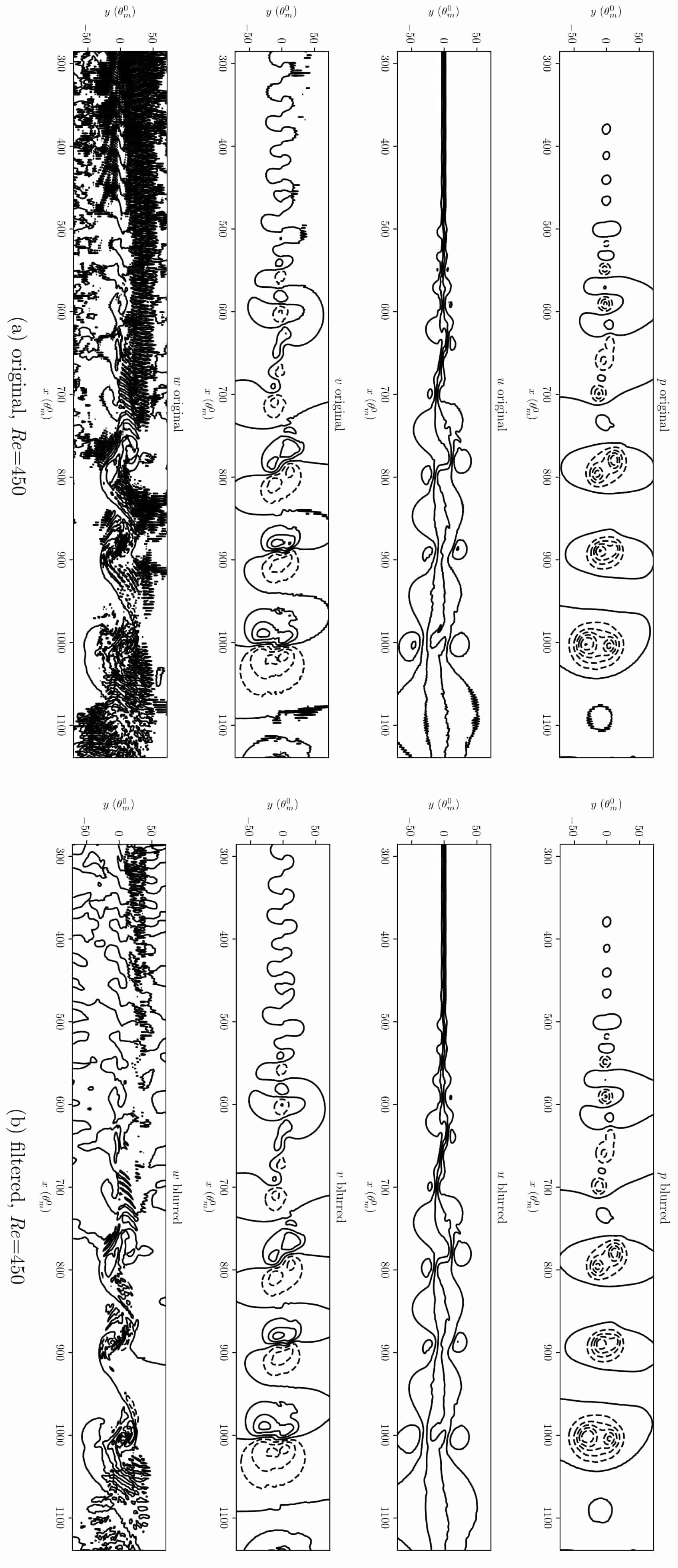}
	\end{subfigure}
	\caption{Comparison between original and Gaussian-blurred data. Small wakes plotted as `noise' of the line plotted in (a,c) obviously decreased in (b,d). Plot is of Re$_m$=450 (a-b) and Re$_m$=800 (c-d), all of them at plane $z=\hat{z}\cdot(L_z/g_z)$, where $\hat{z}=\lfloor g_z/2\rfloor, \hat{z}\in\mathbb{N}$. This $z$-plane was set for 2D visualization of data, which is three-dimensional. Training set and network input are of three dimension as well. Whole simultion result written, which is network input (and therefore plotted here), is truncated region of $y\leq76.0\theta_m^0$.}
	\label{fig:gaussian_line}
\end{figure}

\noindent \ref{simulation}.2 \textit{Mesh grid structure and stretchings}\\
\indent The computational domain size was set, referring to several DNS on mixing layers\cite{rogers1994direct,poinsot1992boundary,moser1993three}. Therefore it is structured Cartesian H-type meshgrid with non-uniform distributions. Here we introduce set $\mathbb{C}=(x,y,z)\in\mathbb{R}^3$ the computational domain which satisfy equations (\ref{xdom})-(\ref{ydom}). The grid is uniformly distributed in streamwise and spanwise direction
\begin{equation}\label{xdom}
x = \Delta x\cdot n \qquad (0\leq n\leq g_x)
\end{equation}
\begin{equation}
z = \Delta z\cdot n \qquad (0\leq n\leq g_z)
\end{equation}
and $n\in\mathbb{Z}$, where $\Delta x=1.52\theta_m^0$ and $\Delta z=1.45\theta_m^0$ following the specification by Stanley and Sarkar\cite{stanley1997simulations}.\\
\indent Although the grid is \textit{coarse} in $x$ and $z$ direction, $y$ has its fine grid spacings. The stretching function by Colonius \textit{et al.}\cite{colonius1993boundary} is not utilized herein despite Stanley\cite{stanley1997simulations} adopted. Considering the feasibility of grid buffer zone in current incompressible flow simulation, current meshgrid structure simply adopted linear stretching:\\
\begin{equation}\label{ydom}
y = \bigcup_{j=1..3} \sum_{i=1}^{j} (n-n_{i-1})\lambda_i
\end{equation}
, which is satisfied $\forall |n|<g_y/2$, $n\in\mathbb{Z}$
with cell growth factors $\lambda_1=0.76\theta_m^0$, $\lambda_2=\theta_m^0$, $\lambda_3=10\theta_m^0$ and cell numbers $n_0 = 0$, $n_1=35$, $n_2=15$, $n_3=29$. Thus it could be said that there exist three distinct regions coupled in $+y$ and $-y$ axes, with three different stretch factors each.\\
\indent Additional to the full domain $\mathbb{C}$, we define truncated region (which is introduced as `box' in \ref{simulation}.1) $\mathbb{B}\subset\mathbb{C}$ for implementation of desired boundary condition. This box is bounded in
\begin{equation}
\mathbb{B} = \{0.25L_x \leq x \leq 0.75L_x\}\cap\{|y| \leq 76\theta_m^0\}
\end{equation}
and the DNS result (which is the actual training set) was written inside $\mathbb{B}$. 
Note that current simulation has $\lceil|y_2|\rceil = 41.6\theta_m^0$. Therefore abrupt increase in $\lambda$ between $n=n_2$ and $n_2+1$, is included in the truncated region. The simulation result showed wake vortices arouse dominantly in range $y<70\theta_m^0$ (see Figure \ref{fig:gaussian_line}).\\

\noindent \ref{simulation}.3 \textit{Measure of lengthscale}\\
\indent Although Stanley and Sarkar adopted vorticity thickness $\delta_\omega$ as their lengthscale, we take momentum thickness $\theta_m$ as lengthscale in simulation and every statistics including data analyses. Main reason for this is well explained in \cite{rogers1994direct} by Rogers and Moser. In their temporally evolving mixing-layer growth of vorticity- and momentum-thickness showed much difference in the stability; the latter being more stabilized. Monkewitz and Huerre\cite{monkewitz1982influence} also stated about growth rate of $\lambda$ in their spatially growing mixing-layer of various boundary conditions. Here the parameter $\lambda=\Delta U/(2\overline{U})$ is a measure of velocity difference, which is also directly correlated with $\delta_\omega/\theta_m$. The ratio is said to be sensitive according to the mean profile given in inflow, by Rogers and Moser\cite{rogers1994direct}. Also the momentum thickness is more general scaling in lots of experiments either, because of its low sensitivity in statistical noise.

\begin{figure}[H]
	\centering
	\includegraphics[scale=0.28]{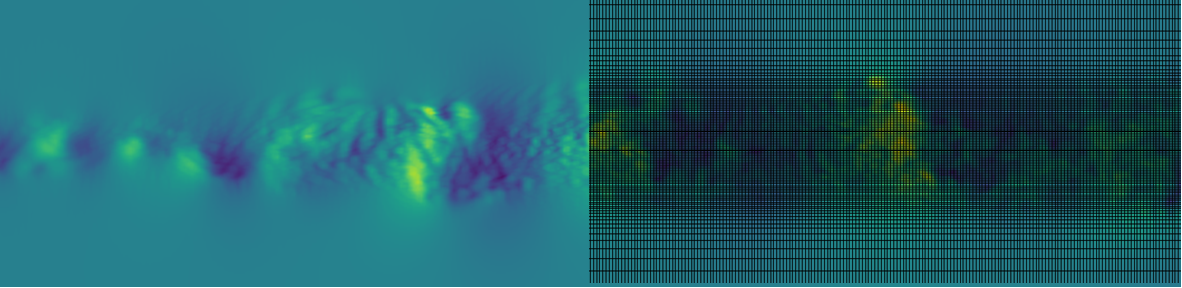}
	\caption{Schematic illustration of meshgrid overlaid on plot of transverse velocity in current simulation result. See Figure \ref{fig:gaussian_line} for details about plot.}
	\label{fig:grid}
\end{figure}

\section{Deep learning}\label{deep_sequence}
As an extension of research by Lee and You\cite{lee2017prediction}, we have predicted turbulent flow features from given those of previous ones in time sequence. Current section will introduce the motivation of flow prediction approach, after briefly introducing backgrounds about artificial neural networks, deep learning, and generative adversarial networks.\\

\noindent \ref{deep_sequence}.0.1 \textit{Short overview of artifical neural network}\\
\indent Neural Network (NN), one of example that comes from reprinting operations of neurons in brain as combination of linear operations of each nodes in network\cite{hassoun1995fundamentals}. Machine learning method by Rumelhart \textit{et al.}\cite{rumelhart1986learning} presented multiple layers of the neuron-like units back-propagating, and showed the importance of `hidden units' for representation of features. Application of back-propagated neural networks to analysis of fluid dynamics, especially turbulence, have been presented \cite{lee1997application, milano2002neural, zhang2015machine, zhang2018machine, lee2018data}.\\
\indent Two fundamentals of NN are network and loss function. A network is a weighted, directed, acyclic graph. Vertices and edges represent neurons and connections between the neurons each, where the weights on the graph represent coefficient for linear combination. Therefore it could be intuitively understood that NN maps from the input array to the output, by liner operation between layers. Loss function is a measure of error, and could possibly be of many kinds even except for the popular ones (e.g. MSE, Least-Absolute- and Least-Squared-Error, Wasserstein Loss\cite{arjovsky2017wasserstein}). Here the error is calculated in between dataset and the output from the graph; and this dataset could be prepared from the prevalent databases (e.g. MNIST, CIFAR), or custom data, and differs according to the purpose of network. NN `evolves' by updating weights of its graph, leading its mapped output to have minimized error. A pair of data can have variety values in error, since the summation of measures can vary according to the adopted loss functions' type. A recent study of San and Maulik\cite{san2018extreme} used fully-connected NN on model reduction of turbulent flows, showing feasibility of NN validated on four-gyre circulation problem in barotropic model.\\
\indent Convolutional Neural Network (CNN) is a deep neural network with `convolution layer', where discrete convolution filtering occurs. This filter kernel is also a kind of weights, which is known to have good performance in extracting target features and their local conjunction. A typical architecture of convolution layer can be represented as feature maps, which is connections of units throughout series of various filters and non-linearity such as Rectified Linear Unit(ReLU). This architecture is intended to have spectacular performance paticularly in two characteristics of data, explained by LeCun \textit{et al.}\cite{lecun2015deep}: One is existance of high correlation among local group of values in an array, and another is irrelevance of location in statistics of data. These features does not indicate image only, but variety of array-like data, including flow data. Miyanawala and Jaiman\cite{miyanawala2017efficient} adopted CNN for analyzing wakes behind bluff-body.\\\mbox{}\\

\noindent \ref{deep_sequence}.0.2 \textit{Network utilized in current paper}\\
\indent Proposed by Goodfellow \textit{et al.}\cite{goodfellow2014generative}, Generative Adversarial Network (GAN) has been showing good performance since 2014. With composed of two major models \textit{generator} and \textit{discriminator}, GAN evolves in adversarial competition between them. The generator creates output and it tries to make it real so that discriminator cannot distinguish; while the discriminator tries to figure out the fake one better. Both of them are improved using backpropagation and dropout by Hinton \textit{et al.}\cite{hinton2012improving}. Beyond the GAN model, Mathieu \textit{et al.}\cite{mathieu2015deep} introduced a methodology adopting gradient difference loss function with least-squared error, amplifying sharpness in predicted data. With modifications to the network of Mathieu, current network predicts three-dimensional flow features when given that previous ones.\\
\indent Deep learning model utilized in current research is generative adversarial network (GAN), but it should be clarified that problem statement is different from those of conventional GANs. Most of the quests given to conventional GANs are to create real-like output. For example, a GAN trained with LSUN bedroom dataset can generate hundreds of figures that looks like bedroom\cite{zhao2016energy}, yet are similar to the actual and typical bedroom, but are not actually the existing places. Therefore the processes cannot be generally esteemed as \textit{solving} but \textit{creating} tasks.\\
\indent Meanwhile GAN is a kind of network with generative model via adversarial process\cite{goodfellow2014generative}. Thus tasks can be more specified, for example providing more conditions on input\cite{isola2017image,liu2016coupled,liu2017unsupervised}. These conditional GANs (cGANs) are differentiated in that they have dependencies on input conditions thus can perceptually be understood that they `modify' conditions, while GANs do not thus start generation from a noisy data. Denoting the data as $x$ the input condition, $y$ the output, and $z$ the noise, cGAN is mapping of $\{x,z\}\mapsto y$\cite{isola2017image} and GAN is of $z\mapsto y$\cite{goodfellow2014generative}. For example, coloring the croquis\cite{isola2017image}, changing a husky into shepherd or corgi\cite{liu2017unsupervised}, converting `The Seine in Argenteuil' by Claude Monet into photo\cite{zhu2017unpaired}. The output data of cGANs are now relatively more constrained but still are creative tasks with no rigid truths.\\
\indent The approach of video prediction has its meaning in that there exist ground truths. The loss functions have dependencies on both. Current network adopting the method of Mathieu, maps $\{x,z\}\mapsto y$. In training process, $x$ includes ground truth $\mathcal{G}$ and generated $G$, and $\mathcal{G}$ is not provided in inference. Therefore cGAN in current research is distinguished from the conventional GANs in that we have specific solution that should be achieved. This can be better explained as a classification model in the point that that there exist a truth for each cells.

\subsection{Preliminaries and motivations}\label{unsteady}
Current subsection illustrates the previous trials and motivation of current paper. The most significant difference is that previous efforts focused on the prediction of unsteady flow in given full DNS results as a sequence of time step, without pixelizations.\\
\indent Images are composition of real values closed and bounded in $[0, 255]\in\mathbb{R}$, red-green-blue(RGB) individually. Thus, those are compact set, and easy to be normalized also. However, determining the upper bound value in DNS results is unachievable unless scanning every digits for numerous numbered cells, making them hard to be normalized.\\
\indent Visualization of simulation results into image files include interpolation and pixelization. This seems to cover the problem stated before, but with inevitable numerical error increment being followed. Moreover the interpolation is linear or second-ordered for most of the cases, flattening all the non-linear turbulent features out. This invoked us of changing learning dataset from plotted images to the raw data.\\

\begin{figure}[H]
	\centering
	\begin{subfigure}{.495\textwidth}
		\includegraphics[width=1\linewidth]{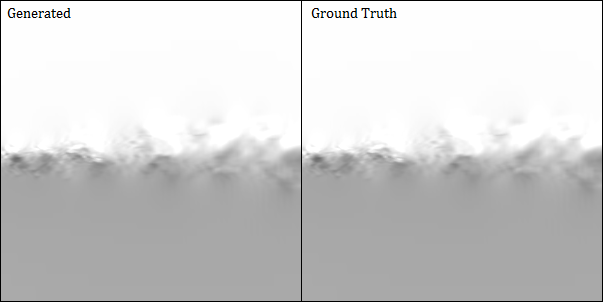}
		\caption{Streamwise velocity $u(t)$}
		\label{fig:splitter_ugen}
	\end{subfigure}
	\begin{subfigure}{.495\textwidth}
		\includegraphics[width=1\linewidth]{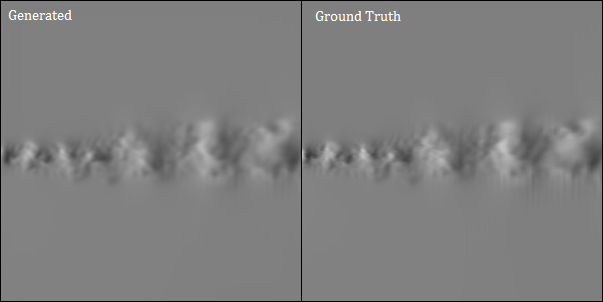}
		\caption{Transverse velocity $v(t)$}
		\label{fig:splitter_vgen}
	\end{subfigure}
	\caption{Plot of velocity data, the predicted ones (\textit{left}) and simulated truths (\textit{right}). Results were predicted using generative adversarial network (GAN), which is modified version of \cite{mathieu2015deep} appropriate for flow data. Note that input images are \textit{normalized with mean velocity} then \textit{pixelized} into black and white (as it could be seen above). Thus it is not \textit{flow data} itself, rather concentrated on the shapes of vortex roller, finiding that GAN successfully learns and predicts fluid motion.}
	\label{fig:splitter_gen_result}
\end{figure}

\noindent \ref{unsteady}.1 \textit{Post-processings of DNS data}\\
\indent Most of GANs, which are actually the subset of CNNs, conventionally use images as their input for reasons explained in section \ref{deep_sequence}.0.1 and \cite{lecun2015deep}. Primitive trials of current subsection also utilized plotted images of flow simulation data. Therefore, the DNS results were plotted in every $14.4\theta_m^0/\overline{U}$ time units after being fully-developed. Those plotted \textit{images} were directly put into the network. Not only for the cylinder flow by Lee and You, current mixing-layer had also been predicted in pixels (see section \ref{previous_2d} to \ref{previous_3d_pixel} and Figure \ref{fig:splitter_gen_result}).\\

\noindent \ref{unsteady}.2 \textit{Motivations to current work}\\
\indent Main purpose of the interpolation and plotting is to overcome \textit{non-uniformity} in cell-centered data. Most of the Newtonian flow simulations have its meshgrid non-uniform, even disregarding the structuredness, because the formations of meshgrid directly affect validity in their solutions. Even though the accurate particular solution to Navier-Stokes equation is given, conventional convolution filter which sweep through it does not consider whether it is uniformly distributed or not. Simply indexed into array, the input data definitely get to have distorted scale. In other words, non-uniformity is a quest must be overcome in utilizing raw data as input to the neural network.\\
\indent Another intention in post-process is to reduce size of data. Discretized into fine meshgrid, DNS results have incomparably larger size than those images utilzed in most of neural networks. Therefore pixelizing the flow field into image includes interpolation and downscaling the file size. However it is definite that approximations including interpolation and resizing give rise to the error from true solution of Navier-Stokes equation. Therefore, current work present modifications in concept of previous works, in the way to overcome the two obstacles. This change in concept focus on the prediction without interpolation or pixelization of raw data, and keep it unprocessed except for truncation indeed.

\section{Prediction of small-eddies\\in 3D turbulent mixing-layer}\label{deep_small}
With same DNS result conducted in section \textbf{\ref{DNS}}, we focus on deblurred prediction with given blurry flow field. Therefore this network is quite similar with those we do in Large-eddy Simulation (LES). The condition provided for GAN of current section is large-eddies, and is required to fill out the small-eddies.\\

\subsection{Data preprocessings}\label{les}
Current project is expecting increments in efficiency and accuracy of turbulent flow modeling. Therefore we preprocess the DNS result acquired from section \ref{simulation} filtering out the large-scaled physical features.\\

\noindent \ref{les}.1 \textit{Gaussian filter}\\
\indent In order to acquire large-scale behaviors it has been general method to put filters on, and these filters are usually variant depending on each simulations\cite{bose2010grid}. Rogers and Moser\cite{rogers1994direct} utilized bivariate Gaussian filter on DNS of mixing-layer result, in order to acquire vortex rollers. This Gaussian filter kernel with its measure of filter width equal to momentum thickness, was bivariate for streamwise($x$) and spanwise($z$) direction since it was to determine regions where rollers or braids are defined.\\
\indent Current filter follows the point of \cite{rogers1994direct,rogers1992three}, which diagnosed large-scale rotations using multivariant Gaussian filter of three-dimension. The width $\Delta$ can be implicitly determined by $\varsigma=k\sigma_o$, after the shape of Gaussian bell is determined when $\sigma_o = \theta_m^0$,
\begin{equation}
N(\varsigma) = 2\lfloor\varsigma + 0.5\rfloor + 1, \quad N\in\mathbb{N}
\end{equation}
\begin{equation}\label{filter}
\Delta_i = L_i \frac{N}{g_i}
\end{equation}
$L_i$ indicates $i$-directional domain size, where $g_i$ denotes mesh grid numbers in indicated computational domain. This width $\Delta_i$ variates in each convolutions because of the non-uniform cell distributions. Width parameter $\varsigma$ is implicitely determined to satisfy
\begin{equation}
\Delta_i(\varsigma) < \delta_\omega^0.
\end{equation}
Thus the DNS result is `\textit{blurred}' using $N\times N\times N$ cubic Gaussian filter kernel. Noting that vorticity thickness $\delta_\omega(x)$ have large varieties having fluctuations larger than those in momentum thickness\cite{rogers1994direct} along the $x$ axis, width $\Delta_i$ does not guarantee a rigid floor in vortices' sizes blurring out, even additional to the non-uniform cell distributions and cell sizes.\\

\noindent \ref{les}.2 \textit{Training set and normalization}\\
\indent Each of training data is composed of gaussian blurred ground truths. Both of them in each single trainig data are of shape $32\times32\times32$ randomly cropped. This random selection of cropping data operates as a kind of input noise while training, thus induces \textit{jittering}, which is a conventional way to prevent overfitting\cite{reed1992regularization}.\\
\indent Every batch into the network are normalized with each of their maximum absolute value among every parameters. More specifically this normalization is independent between location $X(n) = \{x(n),y(n),z(n)\}$ and flow parameter tensor which will be explained from now on. First we put the trainig input as
\begin{equation}
\hat{\mathcal{I}}= \Big\{\bigcup_{n\in\mathbb{B}} \hat{G}(n), \bigcup_{n\in\mathbb{B}} \hat{\mathcal{G}}(n)\Big\},
\end{equation}
where $\mathcal{G} = X(n) \cup I(n)$ and $G = X(n) \cup \tilde{I}(n)$, and $\mathcal{G}$ only exists for training. A tilde indicates the Gaussian-filtered quantity, and a hat indicates raw data which is not normalized. Also it is
\begin{equation}
\hat{\mathcal{I}}= \bigcup_{n\in\mathbb{B}} \hat{G}(n)
\end{equation}
for inference process. Current paper will disregard repeated elaborations on $\mathcal{G}(n)$ for explained processses on $G(n)$ in current section, since the processes are equivalent for each set. $\mathcal{I}$ represents a training batch input as unison on every cell points in truncated region $\mathbb{B}$. $n=(i,j,k)$ enumerates the index of each cell points in computational domain. Here the normalization denominator $c$ can be defined
\begin{equation}\label{floor}
f = \lfloor\hat{I}(n_f)\rfloor \qquad\ s.t.\ n_f = \operatorname*{arg\,min}_{n\in\mathbb{B}}\lfloor\hat{I}(n)\rfloor,
\end{equation}
\begin{equation}\label{denom}
c = \Big\lceil|\hat{I}(n_c)|\Big\rceil - f \quad s.t.\ n_c = \operatorname*{arg\,max}_{n\in\mathbb{B}}\Big\lceil|\hat{I}(n)|\Big\rceil.
\end{equation}
$c$ is unique for each set $\hat{\mathcal{G}}$ or $\hat{G}$. Thus for inference, it is unique for each $\mathcal{I}$-s. Similarly, we normalize $X(n)$ using $f_x = \lfloor\hat{X}(n_{f_x})\rfloor$ such that $n_{f_x} = \operatorname*{arg\,min}_{n\in\mathbb{B}}\lfloor\hat{X}(n)\rfloor$, and $c_x = \lceil|\hat{X}(n_{c_x})|\rceil - f_x $ such that $n_{c_x} = \operatorname*{arg\,max}_{n\in\mathbb{B}}\lceil|\hat{X}(n)|\rceil$. Therefore the normalization is 
\begin{equation}\label{norm}
X(n) = \frac{\hat{X}(n)-f_x}{2c_x}, \qquad I(n) = \frac{\hat{I}(n)-f}{2c}
\end{equation}
so that every $\mathcal{I}$ is a compact set in $[0, 1]$ for all $n\in\mathbb{R}$, and $\mathcal{I}$ becomes the real training batch to the GAN. Denormalization is simply a reverse process of normalization. With declared global variables $c$ and $f$ unique for each batches, are being multiplied and added to the normalized frames and the generated data as well. Thus for the output data $G$ of GAN mapping $\mathcal{I}\mapsto G$, denormalization is to $\hat{G} = 2c\times G+f$ where $c$ and $f$ directly comes from the input $\mathcal{I}$ with equation (\ref{floor}) and (\ref{denom}). Although some limitations exist as stated in section \ref{future_deep}, normalization and denormalization are necessary data processings for better analyses of losses.

\subsection{Deep learning model}\label{deep_gan}
\begin{figure}[H]
	\centering
	\includegraphics[scale=0.4]{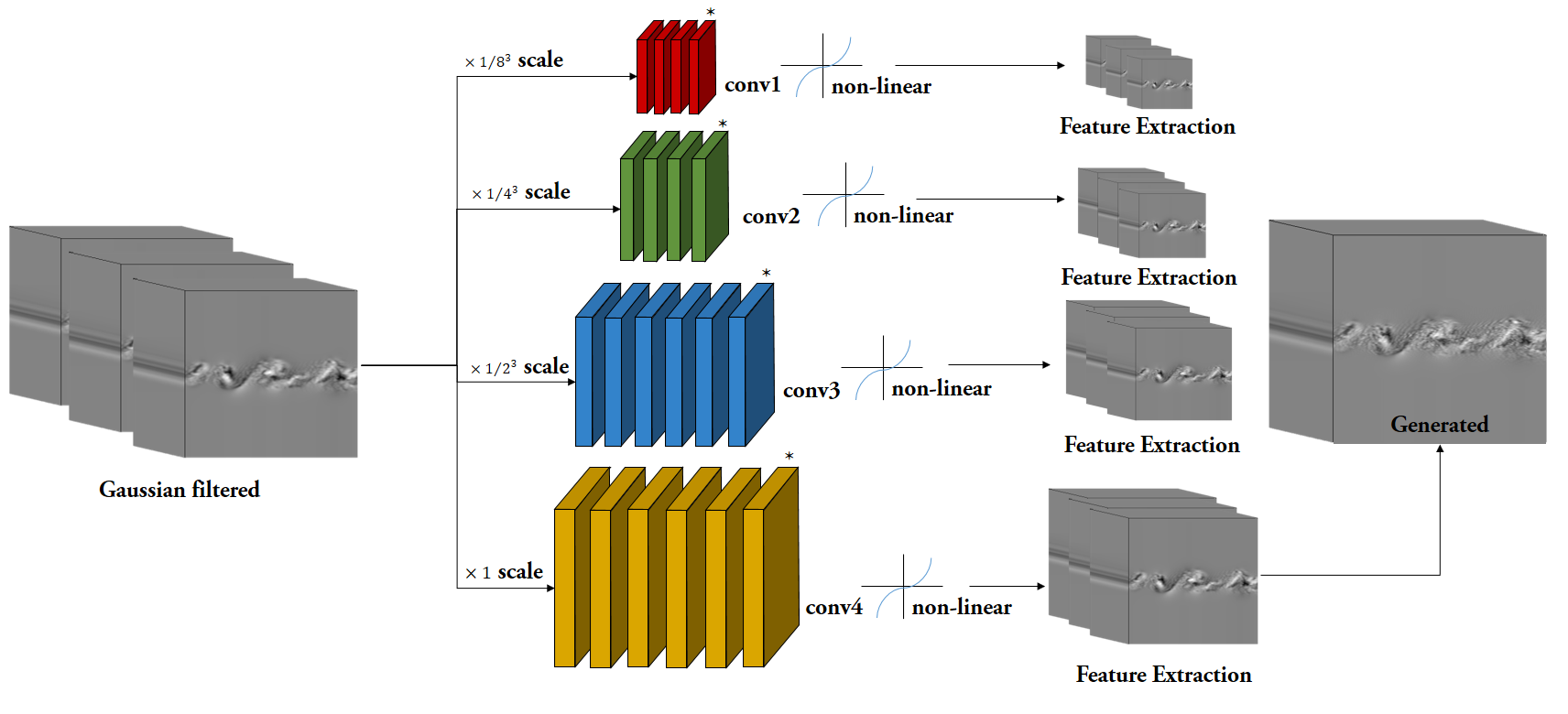} 
	\caption{Schematic diagram of convolution layers in generator network. The feature maps are created for each seven inputs $x, y, z, p, u, v, w$; $\times$($3,128,256,128,1$) for $\times1/8^3$ scale, $\times$($4,128,256,128,1$) for $\times1/4^3$ scale, $\times$($4,128,256,512,256,128,1$) for $\times1/2^3$ scale, $\times$($4,128,256,512,256,128,1$) for $\times1$ scale. Convolution kernel shapes are ($3^3,3^3,3^3,3^3$) for $\times1/8^3$ scale, ($5^3,3^3,3^3,5^3$) for $\times1/4^3$ scale, ($5^3,3^3,3^3,3^3,3^3,5^3$) for $\times1/2^3$ scale, ($7^3,5^3,5^3,5^3,5^3,7^3$) for $\times1$ scale. Cubics $R^3$ for kernel shape denotes shape of $R\times R\times R$. Outputs from the last convolution layers ($*$-s in figure) go through $\tanh$ non-linear function, and the others go for rectified linear unit(ReLU). Input $\mathcal{I}$ is sixth orderd tensor and each orders indicate batch, flow parameter, $x$-, $y$-, $z$-coordinate, and time-step. Here the 3D convolution is conducted for each flow parameters individually.}
	\label{fig:conv}
\end{figure}

\begin{figure}[H]
	\centering
	\includegraphics[scale=0.4]{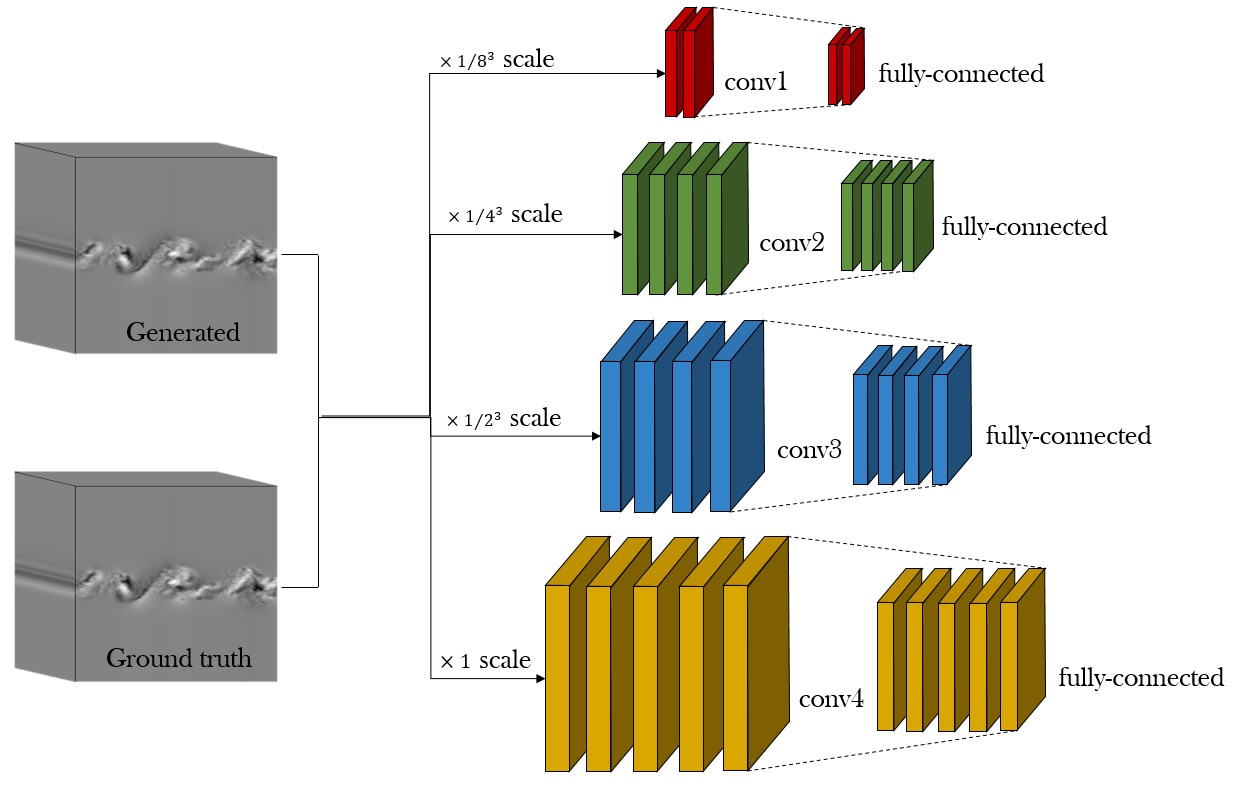} 
	\caption{Schematic diagram of hidden layers in discriminator network. The feature maps are created for each seven inputs $x, y, z, p, u, v, w$; $\times$($1,64$) for $\times1/8^3$ scale, $\times$($1, 64,128, 128$) for $\times1/4^3$ scale, $\times$($1,128,256,256$) for $\times1/2^3$ scale, $\times$($1,128,256,512,128$) for $\times1$ scale. Convolution kernel shapes are ($3^3$) for $\times1/8^3$ scale, ($3^3,3^3,3^3$) for $\times1/4^3$ scale, ($5^3,5^3,5^3$) for $\times1/2^3$ scale, ($7^3,7^3,5^3,5^3$) for $\times1$ scale. The fully-connected layers are of shape ($512,256,1$) for $\times1/8^3$ scale, and ($1024,512,1$) for $\times1/4^3$, $\times1/2^3$, $\times1$ scales.}
	\label{fig:d_conv}
\end{figure}

\noindent\ref{deep_gan}.1 \textit{Generative Adversarial Network}\\
\indent Deep learning model utilized in current research is generative adversarial network (GAN) with flow fields of previous time-steps provided as condition. Schematic illustrations about architectures of generator and discriminator are provided in figures \ref{fig:conv} and \ref{fig:d_conv}. Basically it is generator's task to create desired prediction, and discriminator reinforces generator's performance giving adversarial penalties each other.\\
\indent There are four hidden convolution layers for each of generator and discriminator, and generator has non-linear filtering functions while discriminator has fully-connected layers for feature extraction. Since the training data array shape is $32\times32\times32$ in each cartesian coordiante, and $7$ flow parameters with location, $3$ conditions of previous time-step given additional to the input time-step, thus there are 917,504 cells in each training batches. Convolution filters of width 3, 5, or 7, and were large enough to catch out small-scaled features. Because of large data points in a single set, batch size of less than 10 batches are recommended for efficient learning.\\

\noindent\ref{deep_gan}.2 \textit{Loss functions}\\
\indent Loss functions of generator in current GAN utilize combination of three losses, based on Mathieu \textit{et al.}\cite{mathieu2015deep} further developed by Lee and You\cite{lee2018data}: gradient difference loss $\mathcal{L}_{gdl}$, least squared loss $L_2$, and adversarial loss $\mathcal{L}_{adv}$. Total loss in generator $\mathcal{L}_{gen}$ can be expressed as
\begin{equation}
\mathcal{L}_{gen} = \frac{1}{N}\sum_{n=0}^{N-1} \Big\{\lambda_{gdl}\mathcal{L}_{gdl}^{n} + \lambda_{L_2}\mathcal{L}_{2}^{n} + \lambda_{adv}\mathcal{L}_{adv}^{n} \Big\}
\end{equation}
, where coefficients $\lambda$ are set to be proportional percentage of each losses $\mathcal{L}$, making total sum of them as 1. Total generator's loss is summation of every losses on each convolution layer scales enumerated in $n$. Here, each losses are computed in set $\mathcal{I}\subset\mathbb{B}$ the input flow data as subset of truncated box.\\
\indent Gradient difference loss (GDL) $\mathcal{L}_{gdl}^{n}$ does good performance in generating less blurry output\cite{mathieu2015deep}. Gradient difference loss of three dimension herein can be elaborated as
\begin{equation}\begin{split}
\mathcal{L}_{gdl}^{n} = \sum_{(i,j,k)\in\mathcal{I}} \Big\{||&\mathcal{G}_{i, j, k}^{n} - \mathcal{G}_{i-1, j, k}^{n}|
- |G_{i, j, k}^{n} - G_{i-1, j, k}^{n}||\\
+||&\mathcal{G}_{i, j, k}^{n} - \mathcal{G}_{i, j-1, k}^{n}|
- |G_{i, j, k}^{n} - G_{i, j-1, k}^{n}||\\
+||&\mathcal{G}_{i, j, k}^{n} - \mathcal{G}_{i, j, k-1}^{n}|
- |G_{i, j, k}^{n} - G_{i, j, k-1}^{n}||\Big\}.
\end{split}\end{equation}
The subscripts $i,j,k$ indicate each cartesian grid points of input $\mathcal{I}$ where $\mathcal{G}^{n}$ denotes the unfiltered ground truth and $G^{n}$ the generated prediction. Without taking mean of datasets, GDL takes on the higher ground than mean squared error (MSE) in the sharpness. Since blurry output is one of critical feature which must be get rid of, GDL should be measured and minimized in generator.\\
\indent Least squared loss $\mathcal{L}_2^{n}$ is simply a sum of differences in absolute values squared
\begin{equation}
\mathcal{L}_2^{n} = \sum_{(i,j,k)\in\mathcal{I}} ||\mathcal{G}_{i,j,k}^{n} - G_{i,j,k}^{n}||^2_2.
\end{equation}
\indent Adversarial loss $\mathcal{L}_{adv}^{n}$ is a special case of binary cross entropy (BCE) loss $L_{bce}$. As generator, trying to create real-like fake one, gets loss function as
\begin{equation}
\mathcal{L}_{adv}^{n} = L_{bce}(D^{n}(G^{n}(\mathcal{I})), 1) + W(D^{n}(G^{n}	(\mathcal{I})), 1)
\end{equation}
where
\begin{equation}
L_{bce}(X,Y) = -Y\log(X) - (1-Y)\log(1-X).
\end{equation}
$D^{n}(G^{n}(\mathcal{I}))$ is boolean of discriminator's judgement about $G^{n}(\mathcal{I})$ thus $D^{n}(G^{n}(\mathcal{I}))$ is 1 or 0 if discriminator reads truth or fake respectively.\\

\indent Loss function of discriminator is defined as
\begin{equation}
\mathcal{L}_{dis} = \frac{1}{N}\sum_{n=0}^{N-1} \bigg\{ L_{bce}(D^{n}(G^{n}(\mathcal{I})), 1) + L_{bce}(D^{n}(G^{n}(\mathcal{I})), 0) \bigg\}
\end{equation} 
, and the summation is on each $n$ numbers of scales. \\

\indent For the quantitative evaluation of quality in generated data, we computed peak-signal to noise ratio (PSNR) between ground truth $\mathcal{G}^n$ and prediction $G^n$
\begin{equation}
\text{PSNR} = 10\log\frac{R^2}{\frac{1}{N}\sum_{n=0}^{N-1}|\mathcal{G}^n - G^n|^2},
\end{equation}
where $R=\max \mathcal{G}^n(\mathcal{I})$ is the value of maximum possible value of generated data. For pixelized values $R$ is generally 255, and is 1 for normalized pixels. In current case, we normalize $\hat{\mathcal{I}}$ into $\mathcal{I}$ which is compact set in $[0,1]$, $R$ should be constant as 1. Using the sharpness measure defined by Mathieu\cite{mathieu2015deep}, we have modified differential sharpness measure feasible for three-dimensional data
\begin{equation}
\text{Sharp.diff.} = 10\log\frac{R^2}{\frac{1}{N}\sum_{n=0}^{N-1}\Big(\sum_{\mathcal{I}} | \nabla\mathcal{G}^n-\nabla G^n|\Big)}.
\end{equation}
Note that gradient is on discretized cartesian coordinate $(i,j,k)\in\mathcal{I}$ and defined as $\nabla X = (X_{i,j,k} - X_{i-1,j,k}) + (X_{i,j,k} - X_{i,j-1,k}) + (X_{i,j,k} - X_{i,j,k-1})$.

\section{Conclusion}\label{conclusion}
\begin{figure}[H]
	\centering
	\includegraphics[scale=0.145]{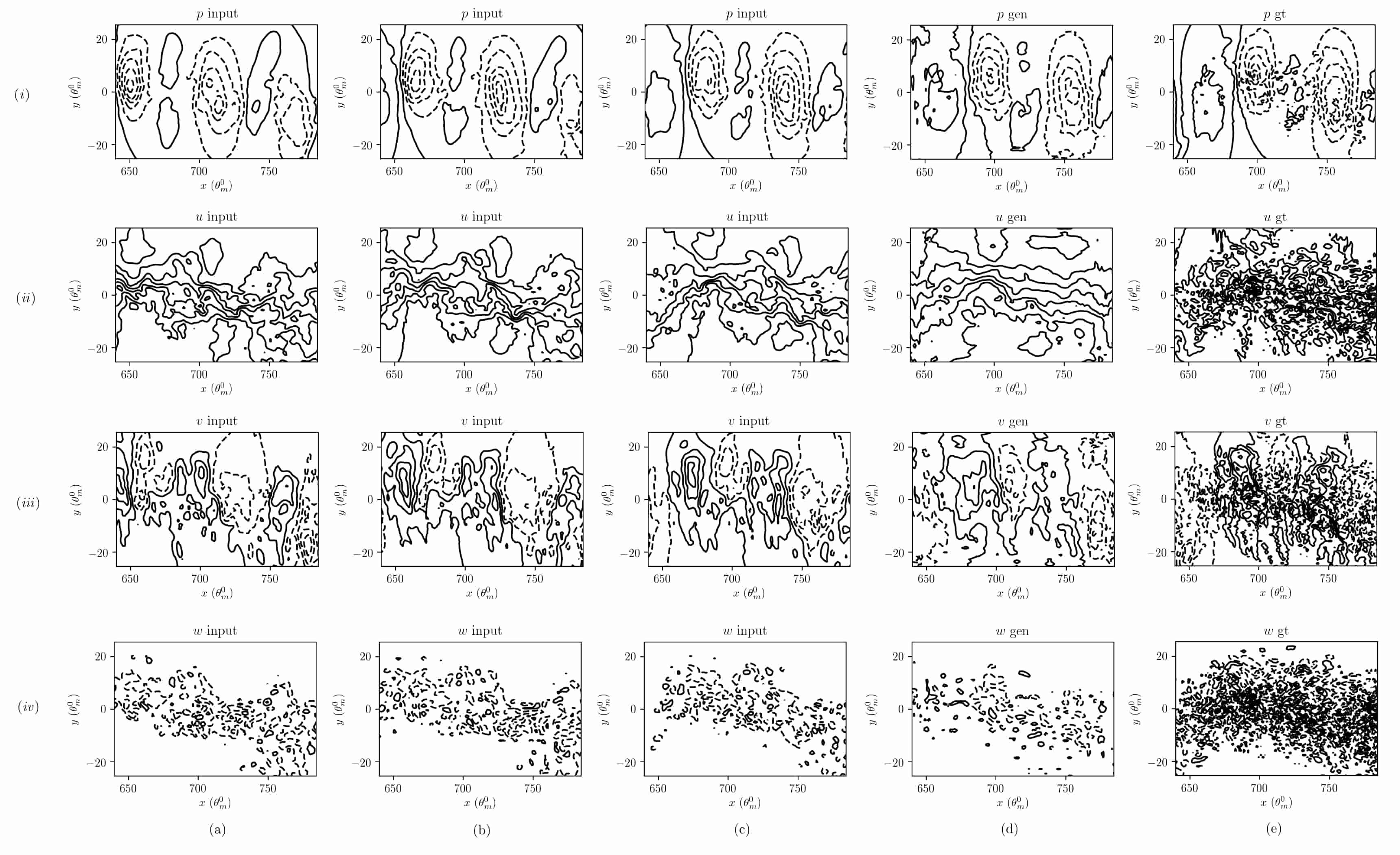}
	\caption{Test result of prediction with model trained for 11000 epoches. PSNR error showed 32.17 and Sharp.diff. error was 25.93 in average throughout parameters. Global loss sum while training the 11000$^{\text{th}}$ step was approximately 535. ($i$) is pressure and the line spaces with 1.2 constant. ($ii$), ($iii$), ($iv$) indicates $x$-, $y$-, $z$-directioned velocity each, with spaces 1.6(minimum value is 4) for $u$, 0.64 for $v$, 0.64 for $w$ also constant. All parameters were put together as a single batch. (a)-(c) are blurred input, (d) is the predicted generation, and (e) shows ground truth which is not blurred DNS data.}
	\label{fig:pred_vel}
\end{figure}

\begin{figure}[H]
	\centering
	\includegraphics[scale=0.145]{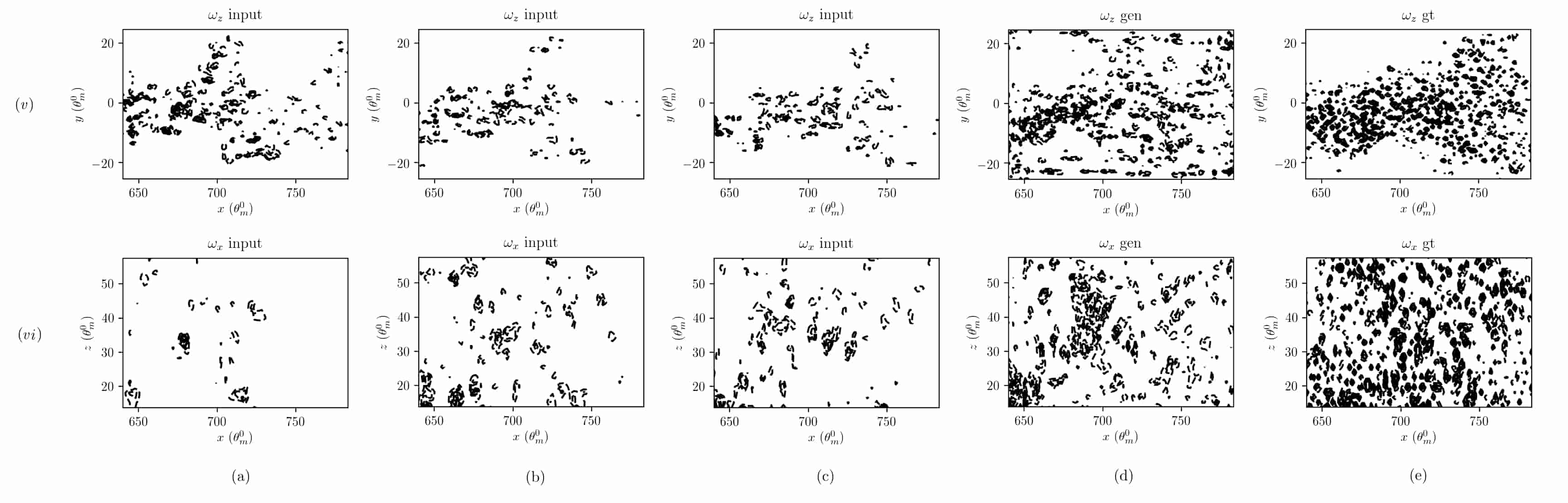}
	\caption{Plot of vorticity ($\lambda_2$-criterion) of test result. ($v$) is plotted on plane $z=\lfloor L_z/2\rfloor\in\mathbb{Z}^+$, and ($vi$) is of plane $y=0$. (a)-(c) are from blurred inputs, (d) is calculated from the prediction, and (e) comes from the gournd truth. Predictions showed more structures of Galilean-invariant vortices\cite{jeong1995identification} with fine scales, but were not sufficient than ground truths.}
	\label{fig:pred_vor}
\end{figure}

Prediction of flow phase with unprocessed mixing-layer data of $p$, $u$, $v$, and $w$ was successful, at Re$_m=800$. Figure \ref{fig:pred_vel} and \ref{fig:pred_vor} show plot of generated predictions (d) with inputs (a)-(c) and ground truths (e). The generated results were succesful in roughly predicting flow features $p, u, v,$ and $w$ with its location and amplitude. Predictions also showed more fine-scaled structures, and visualized vortices in Figure \ref{fig:pred_vor} explains better in possibilities of multi-scaled modeling using deep learning.\\
\indent However those also showed insufficient precision compared to the unfiltered DNS results. This was because of feature extraction mechanism in CNNs. Convolution over data creating abstract features of bulky motions was not enough to predict sufficiently fine-scaled structures. Therefore future developments with modifications of convolution filters, with adoption of other models such as Long-short Term Memory(LSTM) on video representation\cite{srivastava2015unsupervised} or classification models are expected to give better performance.\\
\indent In our DNS results, Q-criterion\cite{hunt1988eddies} was hard to be found while $\lambda_2$\cite{jeong1995identification} was sufficiently found. With different definition of each region for identifying vortices Hunt, Wray \& Moin\cite{hunt1988eddies} defined 
\begin{equation}
Q = \frac{1}{2}\Big[|\mathbf{\Omega}|^2 - |\mathbf{S}|^2\Big] > 0,
\end{equation}
while Jeong and Hussain\cite{jeong1995identification} did
\begin{equation}
\lambda_2(\mathbf{S}^2 + \mathbf{\Omega}^2) < 0,
\end{equation}
where $\mathbf{S} = \frac{1}{2} [\mathbf{J} + \mathbf{J}^\text{T}]$ and $\mathbf{\Omega} = \frac{1}{2} [\mathbf{J} - \mathbf{J}^\text{T}]$ denotes strain rate tensor and vorticity tensor each, which actually are the decomposition of Jacobian matrix $\mathbf{J} = \nabla\mathbf{u} = \mathbf{S} + \mathbf{\Omega}$. $\lambda_2(A)$ is the intermediate eigenvalue of symmetric third-order tensor $A$ of cartesian coordinate. $\lambda_2$ criterion is equivalently a \textit{Galilean-invariant} vortex region. Physical meaning of Q-criterion is the region where rotational component dominates the strain term of fluid. With expression distinguished by Rogers and Moser\cite{rogers1994direct}, Q-criterion is to visualize the `rollers' only, not the `braids'.\\
\indent Absence of Q-criterion was consistent with every data including ground truths. Following the claim by Rogers and Moser, this implies that DNS data itself had almost no region for vortex rollers, and noting the simulation specifications with its cell size $\Delta x = 1.52\theta_m^0$, $\Delta z = 1.45\theta_m^0$, this seems to be trivial results; and will be discussed further in section \ref{future_num}.\\
\indent Current research showed possibilities in deep-learning prediction of flow features with unprocessed 3-dimensional mixing-layer DNS results, without visualizations or interpolations. With further efforts that will be stated in section \ref{future}, we are expecting to predict small-scaled flow features (e.g. turbulent vorticity breakups and dissipations, inertial subrange features), especially in LES and multi-phase flow.\\
\indent Therefore novelty of current research comes from three points: \textit{1) utilization of raw data}, \textit{2) three-dimensional data is being simultaneously processed}, and \textit{3) predictions of small-scaled features with only given large ones}. A single inference process of prediction using current network takes a few minutes only. Regarding conventional methods in modeling small-eddies, current approach shows its good agreement on reduction of computational cost. Also 3D convolutions of current network drives good agreement on 3D feature extractions which is essential in turbulence modeling. Therefore, proposed model, predicting multi-scaled flow features for next time-step, with only given filtered DNS data, is expected to provide adventages in numerous simulations which requires multi-scaled analysis, with high computational cost and accuracy.

\newpage
\section{Future works}\label{future}
\subsection{Numerical simulations}\label{future_num}
\noindent \ref{future_num}.1 \textit{Direct Numerical Simulation}\\
\indent In numerical simulations of current paper, explanations on boundary condition seems like it should be more specified. Even though we have set domain in transverse ($y$) direction large, in order to acquire free in/outflow. Also, still there exist some vaguity in the condition illustrated `free in/outflow'. Current simulation cannot say that there was no effect of free slip $y$-boundaries. Although it has been seen to affect small enough, the support of numerical/mathematical results/proofs seems essential. In other words, reasons for selecting these boundary conditions should be done in future works.\\
\indent Although Stanley and Sarkar\cite{stanley1997simulations} showed close agreement with previous pioneers\cite{rogers1994direct,spencer1971statistical,bell1990development,wygnanski1970two} and thus current DNS followed Stanley's, it had its grid size \textit{coarse} especially in streamwise direction uniform as $\Delta x=1.52\theta_m^0$. This later yielded the width of Gaussian filter proposed in current paper larger than those by Rogers and Moser\cite{rogers1994direct}. Therefore we had the filter width determined by standard deviation in amount of momentum- and vorticity-thickness. More specifications and reasonings should be proposed in future research, otherwise adopting better simuation with fine resolution as input dataset would be preferred.\\

\noindent \ref{future_num}.2 \textit{Filtering of small-eddies}\\
\indent Selection of appropriate width in filter to acquire large-eddies only, in turbulent mixing layer, was one of the most ambiguous question. Even before discussions about eddies, representations of turbulent flow field did not seemed to be appeared generally agreed, as described by Hunt \textit{et al.} \cite{hunt1988eddies}. This ambiguity in distinguishing large- and small-eddy seems to come from relativity in scales of domain. For example, average of absolute sizes of dominant eddies (or simply dominant vortex tubes\cite{hunt1988eddies}) in wakes behind car are much smaller than those in typhoon; and this is trivial because the domain scales are different. In other words, the simulation lengthscale (usually the length term in Reynolds number in CFD) should be one of determinants in saying `how much is the large ones'.\\
\indent  Therefore we have selected the filter width as equation (\ref{filter}) based on the lengthscales blurring out the momentum thickness but smaller than the vorticity thickness. However the filter did not blur out vortices using uniform criterion. This came from non-uniform distributions and the difference in cell size in each coordinates (see section \ref{unsteady}). Thus more scientific quantifications in selection of filter width would be necessary in future works.\\

\subsection{Deep learning model}\label{future_deep}
\noindent \ref{future_deep}.1 \textit{Preprocessing data, and Database}\\
\indent Current research have normalized input data making it closed and bounded in [-1,1]. This assures measures in losses to be feasible to calculate and to be defined. However the normalizing data could also be a kind of threshold, behaving as the ceiling of predicted data phases. For example, let us suppose two batches $\hat{\mathcal{I}}^1\subset[-1,4]$ and $\hat{\mathcal{I}}^2\subset[-5,3]$ (using hat to denote `data not normalized') were imported in a training set. Now what network really gets from those after normalization are $\mathcal{I}^1\subset[-0.25,1]$ and $\mathcal{I}^2\subset[-1,0.6]$; all of them bounded in [-1,1] regardless of what the real values are. Trivially the generated data before denormalization $G(\mathcal{I}^1)$ will be multiplied by 4, and $G(\mathcal{I}^2)$ will be 5 with accordance to each of their denominator, but are highly probable to be bounded in $\hat{G}(\mathcal{I}^1)\subset(-4, 4)$ and $\hat{G}(\mathcal{I}^2)\subset(-5, 5)$ because the training data were always bounded.\\
\indent Normalization is highly recommended in deep learning, especially for vision data, and those pixelized data are clear to be compact set in $[0,255]$ in any cases. With physical raw data however, this \textit{bounded}ness can be a burden in prediction, since it is not clear to say $G$ should be bounded. Thus mathematical approach for quantifying boundedness of generations in raw physical data inputs should be imposed.\\

\noindent \ref{future_deep}.2 \textit{Model structure}\\
\indent As stated before, model of our conditional GAN has some difference in its purpose and dataset. Not like the conventional GANs, we have `answer's (\textit{i.e.} ground truths) for each generated results, thus do not allow creativity. It could be more intuitively accepted that the task of this conditional GAN is for classification in each cells of whole domain.
Therefore trials of having joint with classification model in generator's network is expected to give a possiblility of breakthrough in bounding problem in normalization. With reference on models of Mathieu \textit{et.al.}\cite{mathieu2015deep} and Tran \textit{et.al.}\cite{tran2015learning}, we have found good agreement\cite{lee2018data} in possibility of prediction of flow data using GANs. However in order to overcome further limitations on predictiong 3D raw data, additional research and modifications for developing optimal networks and new models seems to be crucial.

\section{Acknowledgement}
Computing environment of this work is provided by Flow Physics and Engineering Laboratory at Pohang University of Science and Technology. The author would like to thank Sangseung Lee and Donghyun You for giving generous advices and valuable feedbacks.
\end{multicols}

\section{Appendix}\label{appendix}
\subsection{Grid convergence test}\label{g_conv}
\begin{figure}[H]
	\centering
	\includegraphics[scale=0.31]{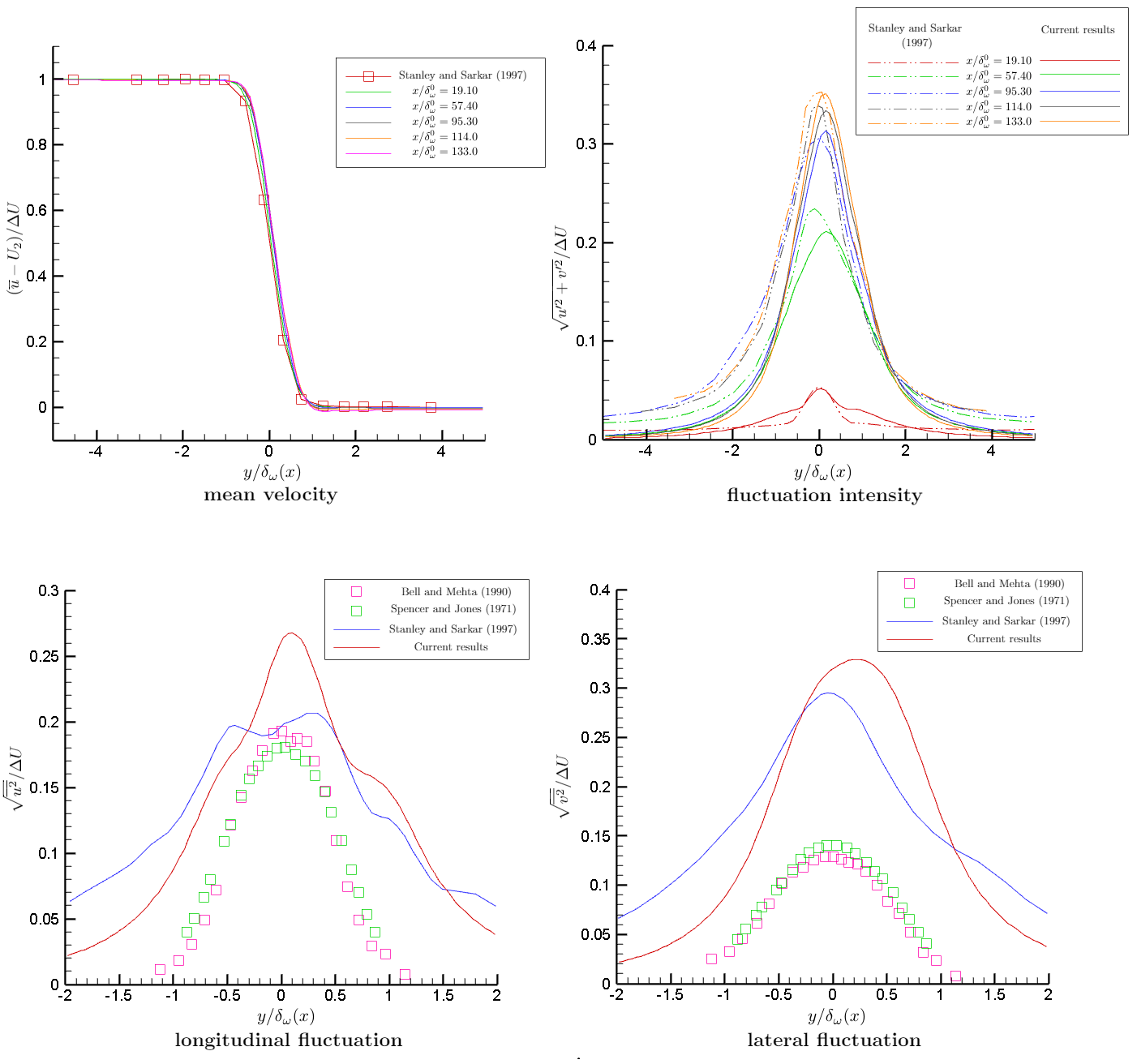} 
	\caption{Velocity profile and fluctuation data comparison with referenced \cite{stanley1997simulations} in 2D result.}
	\label{fig:simres}
\end{figure}

\indent Procedures of assessing robustness in our statistics of data, including grid convergence test result is shown throughout Tables \ref{tab:domain_2d} to \ref{tab:gconv_spec}. The former is of two-dimensional case and is not intended to show grid convergence. Meanwhile the latter is three-dimensionalized simulation from Table \ref{tab:domain_2d}, and is grid convergence test. Table \ref{tab:domain_2d} includes primative domain size with grid refined ($i$-$iii$) and the enlarged domain ($iv$-$vi$). Time-averaged fluctuation intensities and mean velocity profile were analyzed throughout simulations on these varieties of domain and meshgrids. Enlarged domain ($v$) showed good agreement with compared to those by Stanley and Sarkar\cite{stanley1997simulations} shown in Figure \ref{fig:simres}.
\begin{table}[H]
	\centering
	\begin{tabular}{ c | c | c c c c}
		option & \code{enum} &   $L_x$ &	$L_y$ &	$g_x$ &	$g_y$\\\hline\hline
		$g_y\times1$ &  &	570 &	1140 &	375 &	158\\\hline
		$g_y\times2$ & ($i$) &	570 &	1140 &	375 &	316\\
		$g_y\times3$ & ($ii$) &	570 &	1140 &	375 &	474\\
		$g_y\times4$ & ($iii$) &	570 &	1140 &	375 &	632\\\hline
		$L_x\times1.5$ & ($iv$) &	855 &	1140 &	563 &	158\\
		$L_x\times2$ & ($v$) &	1140 &	1140 &	750 &	158\\
		$L_x\times1.5;\ g_y\times2$ & ($vi$) &	855 &	1140 &	563 &	158\\\hline
	\end{tabular}
	\caption{Modifications in domain size and grid distributions in 2D simulation\cite{stanley1997simulations}. Assessed on Reynolds number of \textit{Re}=180. See Figure \ref{fig:simres} for comparisons of $(v)$ with several preliminary simulations \cite{stanley1997simulations,spencer1971statistical,bell1990development}. $L_i$ indicates $i$-directional domain size, where $g_i$ denotes grid numbers. \label{tab:domain_2d}}
\end{table}
\begin{table}[H]
	\centering
	\begin{tabular}{ c | c | c c c | c c c | c}
		option & \code{enum} &   $L_x$ &	$L_y$ & $L_z$ & $g_x$ &	$g_y$ &	$g_z$ & \\\hline\hline
		$g_y\times0.5$ & ($v$-a) &	1140 &	1140 & 71.25 & 	750 &	79 & 49 & half grid\\
		$g_y\times1$ & ($v$) &	1140 &	1140 & 71.25 & 	750 &	158 & 49 & conventional\\
		$g_y\times2$ & ($v$-b) &	1140 &	1140 & 71.25 & 	750 &	316 & 49 & double grid\\\hline
	\end{tabular}
	\caption{Specification of grid convergence test. Assessed on \textit{Re}=800, amongst simulated Reynolds number of \textit{Re=}450, 800, and 1200. See Figure \ref{fig:gridconv} for analysis on velocity profile and fluctuation intensity during grid convergence. $L_i$ indicates $i$-directional domain size, where $g_i$ denotes grid numbers. \label{tab:gconv_spec}}
\end{table}

Data was extracted from the truncated domain within the overall enlarged one.\\
The domain was trunctated based on the position of vortex rollup. Reference explained that rollup was evoked within $x=50\delta_w$ to $x=80\delta_w$. Current simulation had its truncation with $\hat{x}$ direction translation of $95\delta_w$.\\

\begin{table}[H]
	\centering
	\begin{tabular}{ c | c c c c c | c}\hline
		$Re_m$ & \ & $\frac{\sqrt{\overline{u'^2}}}{\Delta U}$ & \ & $\frac{\sqrt{\overline{v'^2}}}{\Delta U}$ & \ & Reference\\\hline
		800 & \ & 0.16 & \ & 0.13 & \ & Rogers and Moser (1994) \cite{rogers1994direct} \\
		- & \ & 0.176 & \ & 0.138 & \ & Wygnanski and Fiedler (1970) \cite{wygnanski1970two} \\
		- & \ & 0.19 & \ & 0.12 & \ & Spencer and Jones (1971) \cite{spencer1971statistical} \\
		$\approx450$ & \ & 0.18 & \ & 0.14 & \ & Bell and Mehta (1990) \cite{bell1990development} \\
		180 & \ & 0.20 & \ & 0.29 & \ & Stanley and Sarkar (1997) \cite{stanley1997simulations} \\ \hline
		180 & \ & 0.1588 & \ & 0.1835 & \ & Current, half grid ($v$-a)\\
		180 & \ & 0.1576 & \ & 0.1882 & \ & Current, original ($v$)\\
		180 & \ & 0.1529 & \ & 0.1899 & \ & Current, double grid ($v$-b)\\\hline
	\end{tabular}
	\caption{Comparison on centerline($\ell$) fluctuation intensities with referenced papers of investigation on free shear layer using simulation\cite{stanley1997simulations,rogers1994direct} and experiment\cite{spencer1971statistical,bell1990development,wygnanski1970two}, with dimensionality of two\cite{stanley1997simulations,wygnanski1970two} and three\cite{rogers1994direct,spencer1971statistical,bell1990development}, on various Reynolds number of momentum thickness. \label{tab:fluct}}
	
\end{table}

\noindent The centerline indicates line $\ell$ herein, which is determied in set  $(x,y,z)\in\mathbb{C}$ the computational domain, bounded in
\begin{equation}
\chi = \{(x,y,z)|\text{ } 0.25L_x \leq x \leq 0.75L_x\}\subset\mathbb{B}
\end{equation}
, thus $\ell$ having finite length is
\begin{equation}
\ell = \chi\cap\{y=0\}\cap\{z= 0.5L_z\}.
\end{equation}
Here the fluctuations along $\ell$ roughly reach the 3D references, yet are insufficient to that of Stanley and Sarkar. This is because of difference in dimensionality. Since current simulation is of three-dimensional, there is much more energy than 2D case. This is because of vortex stretching (which is not achieved in two-dimensions), and can be explained looking into the vorticity equation; this is described in section \ref{kolmo}. It is shown that
\[-\overline{u_i'u_j'}\cdot\frac{\partial U_i}{\partial x_j}= \nu\overline{\zeta_i'\zeta_j'} = \epsilon\]
, where $\overline{u_i'u_j'}/2$ is mean energy and $\overline{\zeta_i'\zeta_j'}/2$ enstrophy. Thus production and dissipation of eddy kinetic energy balace should be matched\cite{james1989lecture} for overall energy level to be acquired. In other words, 3D simulation has more vortex stretched than 2D which trivially increases enstrophy $\overline{\zeta_i'\zeta_j'}/2$, thus the mean energy $\overline{u_i'u_j'}/2$ is higher in 3D for balance.\\

\subsection{Previous Works}\label{previous}
Current section illustrates timeline of our previous efforts on flow feature predictions using deep learning. Primitive trial was to predict using  visulaized ones (images plotted 2D image from 3D simulation by Laizet \textit{et al.}\cite{laizet2010direct}). Next approach we used 2D simulation by Stanley and Sarkar\cite{stanley1997simulations}. Finding out this mixing layer simulation was well-reconstructed and predicted also, third step was to expand dimension into 3D. The 3D domain size and boundary conditions were selected referring to \cite{rogers1994direct,poinsot1992boundary,moser1993three}. Last effort, which became one of motivations in current paper was to utilize raw-data instead of pixelized images. Thus we had confronted some difficulties in directly utilizing heavy, non-uniform DNS result into deep learning input. Current paper intended to develop an approach that deals with these obstacles.

\subsubsection{\textnormal{\textit{Pixel-based prediction of 3D mixing-layer downstream thick splitter plate}\cite{laizet2010direct}}}
\begin{figure}[H]
	\centering
	\begin{subfigure}{.32\textwidth}
		\centering
		\includegraphics[width=1\linewidth]{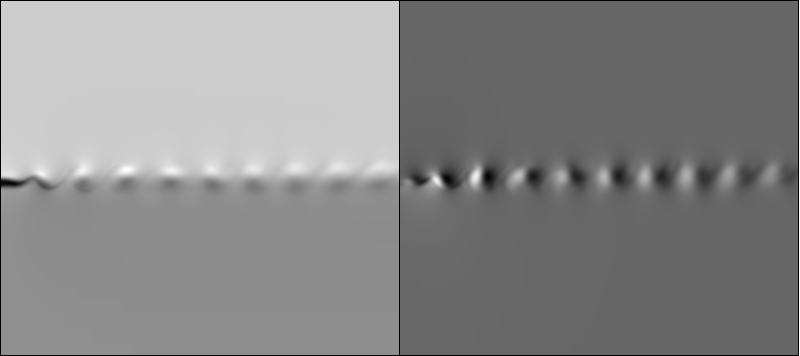}
		\caption{\textit{Re}=200}
		\label{fig:splitter200}
	\end{subfigure}
	\begin{subfigure}{.32\textwidth}
		\centering
		\includegraphics[width=1\linewidth]{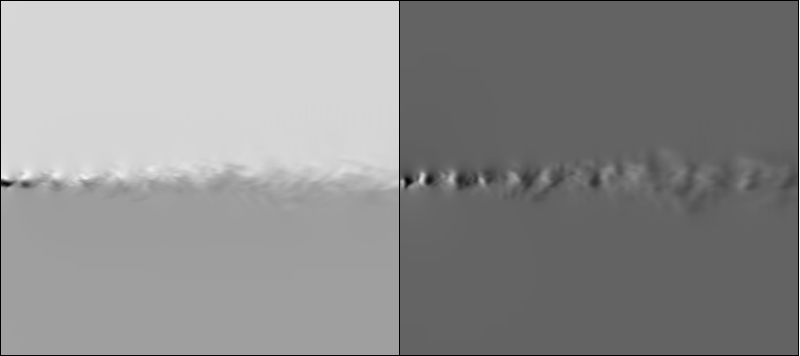}
		\caption{\textit{Re}=800}
		\label{fig:splitter800}
	\end{subfigure}
	\begin{subfigure}{.32\textwidth}
		\centering
		\includegraphics[width=1\linewidth]{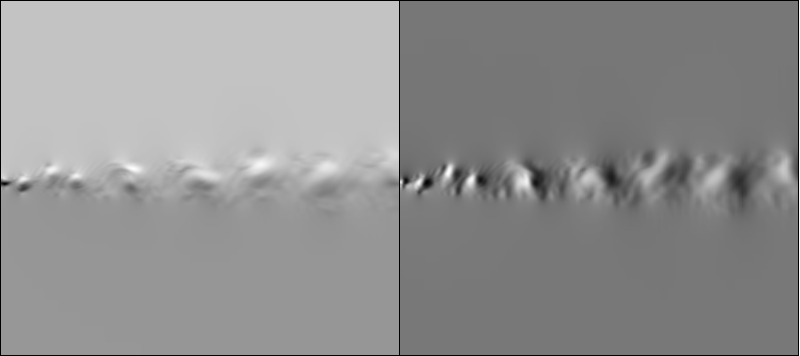}
		\caption{\textit{Re}=1000}
		\label{fig:splitter1000}
	\end{subfigure}

	\caption{Plot of simulation result, reconstructed from 3D mixing layer downstream a thick splitter plate\cite{laizet2010direct} with a blunt trailing edge (BTE). Specifications followed were along with descriptions by Laizet \textit{et al} (2010). Flow images above were plotted in quasi-continuous contour levels, and these were put into neural network after preprocessing. Left hand side of each figures (a)-(c) are of streamwise velocity $u(t)$, and the right hand side are of transverse velocity $v(t)$.}
	\label{fig:splitter_plot}
\end{figure}

As the most primitive approach to prediction, we had followed some DNS of mixing-layer and build up database for our learning model. First DNS we had referenced is those after splitter plate with blunt trailing edge, by Laizet \textit{et al.}\cite{laizet2010direct}. The plotted simulation result in \textbf{Figure \ref{fig:splitter_plot}} is trials with coarser meshgrid than those of Laizet. Reynolds number were on \textit{Re}=200, 800 and 1000. Plotted images of those simulation results show more turbulent in higher reynolds number, with more vorticity breakups into smaller eddies. Because of high computational power requirements, other papers of DNS were surveyed.
\subsubsection{\textnormal{\textit{Pixel-based prediction of 2D mixing-layer}\cite{stanley1997simulations}}}\label{previous_2d}
Spatially growing mixing-layer by Stanley and Sarkar\cite{stanley1997simulations} was adopted as next trial. This simulation has two major differences; absence of splitter plate, and reduction of dimension into 2D. Therefore this simulation exactly followed illustrations by Stanley and Sarkar, except for numerical scheme. Stanley and Sarkar utilized third-order Runge-Kutta scheme by Willamson\cite{williamson1980low} which is a step higher than those of ours. Therefore this simulation was intended not for prediction but to roughly check out feasibility of our solver in simulation by Staney and Sarkar. The result was esteemed to be proper, thus we moved on to expanding our dimension to three, in order to observe turbulence and hopefully rib structures.
\subsubsection{\textnormal{\textit{Pixel-based prediction of 3D mixing-layer}}}\label{previous_3d_pixel}
With finding out good agreement in possibility of prediction in flow features using GANs, we had expanded domain of simulation data into 3D. More than 3-dimensionalization of \ref{previous_2d}, current version has a difference; the boundary conditions and domain size are set, according to \cite{rogers1994direct,poinsot1992boundary,moser1993three}. Poinsot and Lele\cite{poinsot1992boundary} simulated viscous compressible mixing-layer flow with non-reacting slip boundary conditions. Moser and Rogers \cite{moser1993three} and Rogers and Moser\cite{rogers1994direct} did temporally developing mixing-layer, with Blasius velocity profile. Despite difference in velocity profile, adopted thickness proportional to those, with boundary conditions followed gave fine agreements on references as shown on \textbf{Table \ref{tab:fluct}}.

\subsubsection{\textnormal{\textit{Raw data-driven prediction of 3D mixing-layer}}}\label{previous_raw}
Tran \textit{et al.} introduced 3D Convolution Neural Network\cite{tran2015learning} of which they clearly state difference with 2D convolution on multiple frames. He put emphasis on the good feasibility of the 3D ConvNet's learning on spatiotemporal feature. Even disreagrding how suitable the 3D convolution is herein, we agree with the point that raw flow data can show much more physical features than those in pixels. Thus without processings, we used raw 3D data as input. The network was modified into three-dimensional convolution and fully-connected layers.\\
\indent However the learning could not even be started because of the data size. Result CFD data are incomparably larger than the conventional ones used in deep-learning. DNS data which are over 1 GB each without any preprocesses, did not seem to give us expectations that much. Even we have truncated and rearranged data (e.g. neighboring-cell, cell structure data) into nothing but flow parameters ($u, v, w, p$) in each location ($x, y, z$) was still insufficient with its size having around 400 MB. Even preprocessing training data, which is randomly cropping the data and save 5 million binary files, was expected to take years of calculation with even 16-threads of CPUs multiprocessed (this never takes more than a week in general, when it comes to a single CPU processing image files). Nevertheless, there were two major obstacles we had confronted: (\textit{i}) non-uniformity in raw CFD data (\textit{ii}) calculation cost followed by huge data size.\\
\indent These two points were incompatible to be solved, because we needed more input parameters in order to overcome (\textit{i}), making it heavier at the same time. Therefore, as an approach to make a small procedure to solve this, we have modified our concept a bit, concentrating on problem solvings, thus we propose current research.

\subsection{Plots and Figures}\label{plots}


\begin{figure}[H]
	\centering
	\includegraphics[scale=0.31]{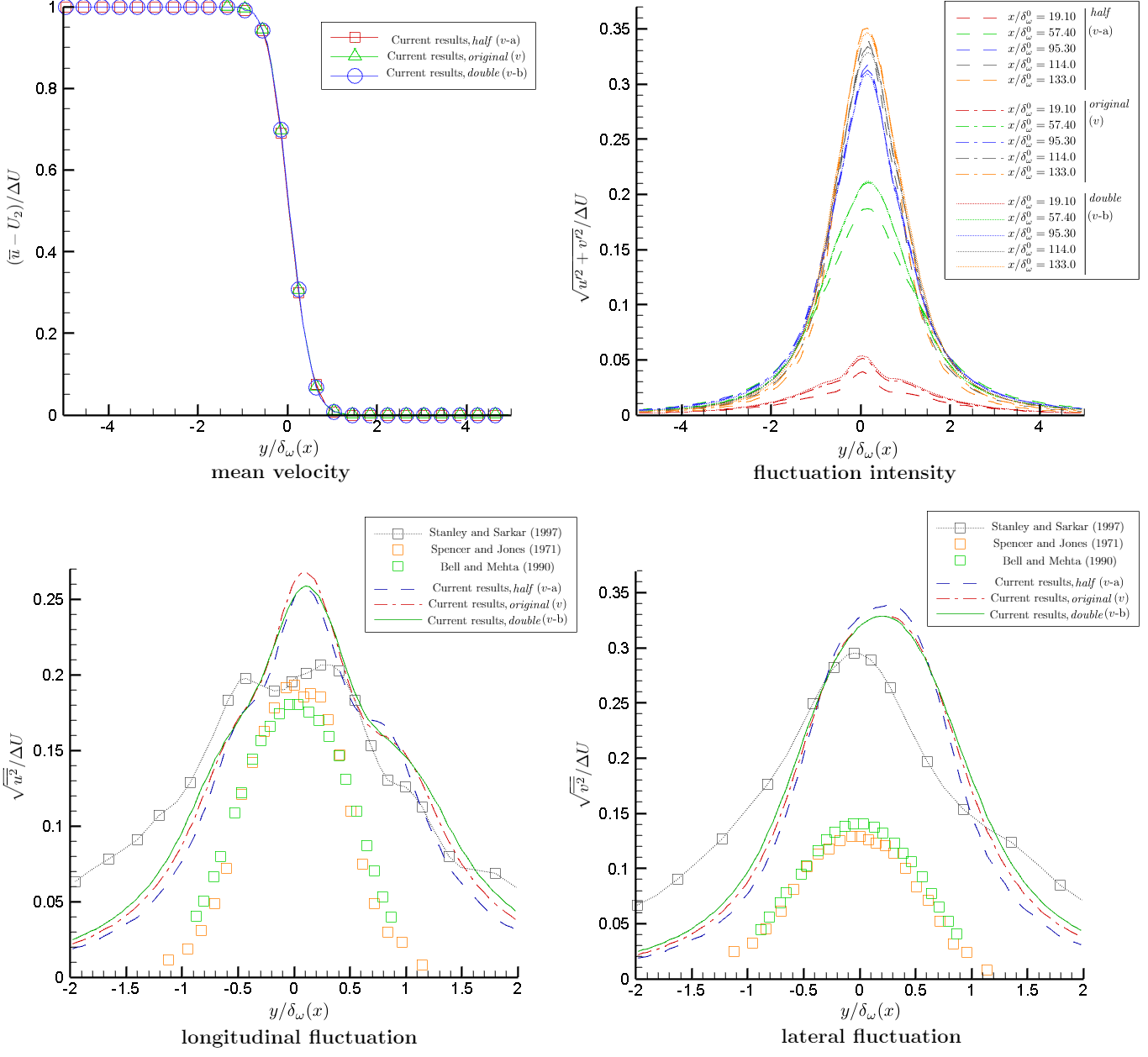} 
	\caption{Grid convergence test result, considering vorticity thickness at each streamwise locations and spatial evolutions. See Table \ref{tab:gconv_spec} for enumerations of each cases. }
	\label{fig:gridconv}
\end{figure}
\begin{figure}[H]
	\begin{subfigure}{.33\textwidth}
		\centering
		\includegraphics[width=1\linewidth]{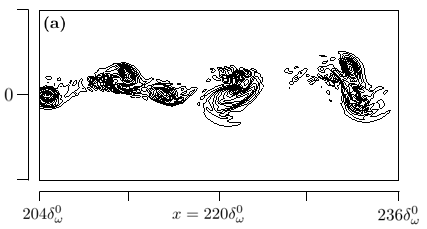}
	\end{subfigure}
	\begin{subfigure}{.33\textwidth}
		\centering
		\includegraphics[width=1\linewidth]{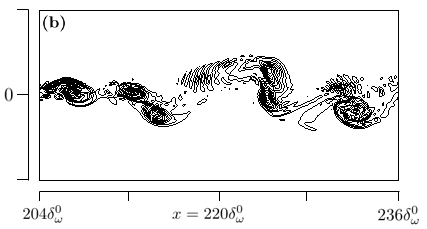}
	\end{subfigure}
	\begin{subfigure}{.33\textwidth}
		\centering
		\includegraphics[width=1\linewidth]{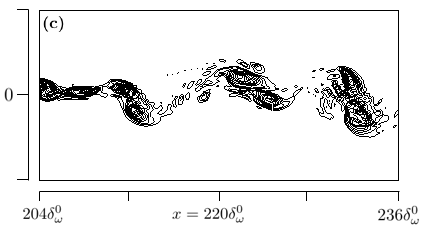}
	\end{subfigure}
	\caption{Line contours of spanwise vorticity in $z=0$ plane of $Re=450$. Figures (a)-(c) shows good agreement with vortices starting to pair at $x=212.5 \delta_\omega^0$, merge at $x=223.0 \delta_\omega^0$, and almost completely paired at $x=231.5 \delta_\omega^0$. First two figures (a) and (b) are plotted when the fluid passed whole domain over 10-th order, while (c) is of 20-th order. Vortex pairing appeared in random sequence, repeatedly.}
	\label{fig:pairingthree}
\end{figure}
\begin{figure}[H]
	\centering
	\begin{subfigure}{.8\textwidth}
		\centering
		\includegraphics[width=1\linewidth]{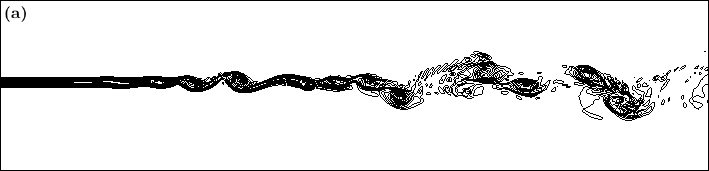}
	\end{subfigure}
	\begin{subfigure}{.8\textwidth}
		\centering
		\includegraphics[width=1\linewidth]{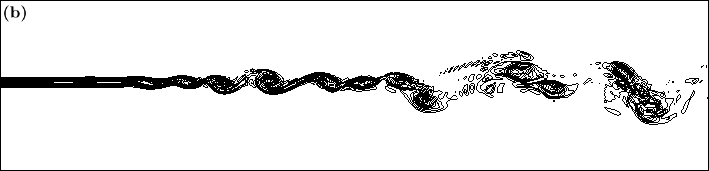}
	\end{subfigure}
	\begin{subfigure}{.8\textwidth}
		\centering
		\includegraphics[width=1\linewidth]{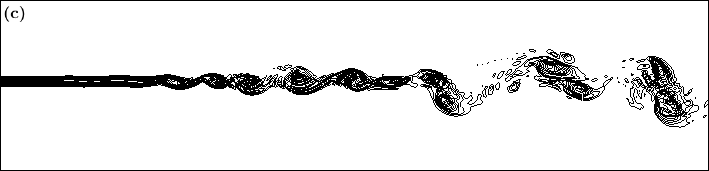}
	\end{subfigure}
	\begin{subfigure}{.8\textwidth}
		\centering
		\includegraphics[width=1\linewidth]{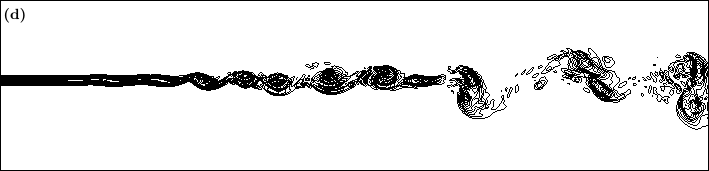}`
	\end{subfigure}
	\caption{Sequence of spanwise vortex pairing in $z=0$ plane of $Re=450$. Timestep between each figures (a)-(d) is static with non-dimensionalized time of 28.80. Flow passed the whole domain (which is double of plotted domain) in 20-th orders, thus could be esteemed as fully developed.}
	\label{fig:pairingsequence}
\end{figure}


\subsection{Brief review of Kolmogorov theory}\label{kolmo}
\indent Since current simulation is of three-dimensional, there is much more energy than 2D case. This is because of vortex stretching (which is not achieved in two-dimensions), and can be explained looking into the vorticity equation.
\begin{equation}\label{vorticity_eqn}
\frac{\partial \zeta_i}{\partial t} + \frac{\partial \zeta_iu_j}{\partial x_j} = \zeta_je_{ij} + \nu\frac{\partial^2 \zeta_i}{\partial x_j\partial x_j}
\end{equation}
, where $e_{ij}$ denotes deformation rate tensor (symmetric part of velocity gradient tensor\cite{james1989lecture}),
\[e_{ij} = \frac{1}{2}(\frac{\partial u_i}{\partial x_j} + \frac{\partial u_j}{\partial x_i})\]
This deformation rate tensor indicates \textit{turning} if $i\ne j$, and \textit{stretching} in $i=j$ case, from r.h.s. of equation (\ref{vorticity_eqn}). Also, $e_{ij}$ vanishes when we lose dimensionality in $i$, making it 2D as follows:\\
\[\frac{\partial \zeta}{\partial t} + \frac{\partial \zeta u_j}{\partial x_j} = \nu\frac{\partial^2 \zeta}{\partial x_j\partial x_j}\]
Therefore, starting with the energy equation, driven from Newtonian Navier-Stokes eqn.(\ref{ns_equation}),
\begin{equation}
\frac{\partial}{\partial t}\frac{1}{2}\overline{u_iu_i} + \frac{\partial}{\partial x_j}(\overline{u_iu_i}) \overline{u_j} = -\frac{\partial}{\partial x_j}(\frac{1}{\rho}\overline{u_jp} + \frac{1}{2}\overline{u_iu_iu_j}) + 2\nu\frac{\partial}{\partial x_j}\overline{u_ie_{ij}} - \overline{u_iu_j}\cdot\overline{e_{ij}} - 2\nu\overline{e_{ij}e_{ij}}.
\end{equation}
Exclusion of flux-divergence and considering homogeneous terms only, it is found that
\[-\overline{u_iu_j}\cdot\overline{e_{ij}} = \nu\overline{\zeta_i\zeta_j}.\]
Although the l.h.s. term is not so clear to be positive right now, it is phenomenologically driven that $\overline{e_{ij}}=\partial U_i/\partial x_j > 0$ in \cite{james1989lecture}. Now the following equation tells that,
\[-\overline{u_iu_j}\cdot\frac{\partial U_i}{\partial x_j}= \nu\overline{\zeta_i\zeta_j} = \epsilon\]
the production and dissipation of eddy kinetic energy balace should be matched\cite{james1989lecture} for overall energy level to be acquired. Therefore directly from r.h.s. of equation above, it is acquired that
\begin{equation}\label{balance}
\overline{\zeta_i\zeta_i} = \frac{\epsilon}{\nu} .
\end{equation}
This equation as a kind of relation of large-scale($\epsilon$) and small-scale($\nu$) parameters; the enstrophy (dissipation term of kinetic energy in fluid flow) increases if $\epsilon$ \textit{increases} or $\nu$ \textit{decreases}. Now for the enstrophy
\begin{equation}
\overline{\zeta_i\zeta_j\frac{\partial u_i}{\partial x_j}} = \nu\overline{\frac{\partial \zeta_i}{\partial x_j}\cdot\frac{\partial\zeta_i}{\partial x_j}}
\end{equation}
and the equation (\ref{balance}), we get Kolmogorov microscale $\eta$ as follows:
\begin{equation}\begin{split}
\sqrt{\frac{\epsilon}{\nu}}\cdot\sqrt{\frac{\epsilon}{\nu}}\cdot\sqrt{\frac{\epsilon}{\nu}} &= \nu\frac{\sqrt{\frac{\epsilon}{\nu}}}{\eta}\cdot\frac{\sqrt{\frac{\epsilon}{\nu}}}{\eta}\\
\Rightarrow \eta &= (\frac{\nu^3}{\epsilon}\large)^{1/4}.
\nonumber\end{split}\end{equation}
Or in current case of simulation, this could also be expressed in form
\begin{equation}
\frac{\eta}{\theta_m^0} = Re^{-3/4},
\end{equation}
by definition of Reynolds number.

\bibliographystyle{unsrt}
\bibliography{reference}

\end{document}